\documentclass[aps,prd,nofootinbib,twocolumn,superscriptaddress]{revtex4}

\usepackage[dvips]{graphicx}

\newcommand{\beq}{\begin{equation}}
\newcommand{\eeq}{\end{equation}}

\newcommand{\Eq}[1]{Eq.~(\ref{#1})}

\newcommand{\bea}{\begin{eqnarray}}
\newcommand{\eea}{\end{eqnarray}}

\newcommand{\gsim}{\lower.7ex\hbox{$\;\stackrel{\textstyle>}{\sim}\;$}}
\newcommand{\lsim}{\lower.7ex\hbox{$\;\stackrel{\textstyle<}{\sim}\;$}}

\begin{document}

\title{Implications of the effective axial-vector coupling of gluon
\vspace{0.1cm} \\
 on top-quark charge asymmetry at the LHC}

\author{Emidio Gabrielli}
\email[]{emidio.gabrielli@cern.ch}
\author{Antonio Racioppi}
\email[]{antonio.racioppi@kbfi.ee}
\author{Martti Raidal}
\email[]{martti.raidal@cern.ch}
\affiliation{NICPB, Ravala 10, 10143 Tallinn, Estonia}

\begin{abstract}
We study different top quark charge asymmetries and the variation of $t\bar t$ total cross section induced by the effective axial-vector coupling of gluon 
in the LHC experiments. We show that rapidity cut-dependent asymmetries are more sensitive to the new physics than the independent ones.
We also study the dependence of  the asymmetries and variations of  total $t\bar t$ cross sections on the invariant mass of $t\bar t$ system
and show that  it would be necessary to measure those quantities as functions of $m_{tt}$ at the LHC.
In the context of considered new physics scenario, 7~TeV LHC has enough sensitivity either to confirm the Tevatron top asymmetry anomaly
 or to rule it out.  In the latter case the LHC is able to put stringent constraint on the new physics scale $\Lambda$ in this framework.

\end{abstract}

\maketitle

\section{Introduction}

The more than $3\sigma$ excess in  top quark charge asymmetry  observed both by the CDF~\cite{Aaltonen:2011kc} and, more recently, by the
D0~\cite{Abazov:2011rq}  experiments at the Tevatron
compared to the standard model (SM) predictions~\cite{Kuhn:1998jr,Kuhn:1998kw,Bowen:2005ap,Antunano:2007da} has triggered
numerous theoretical and experimental studies of top quark production at hadron colliders.
An intriguing  property of the measured asymmetry is that it increases with the $t\bar t$ invariant mass $m_{tt}.$
At the same time the measured $t\bar t$ production cross section is consistent, within experimental errors,
 with the SM predictions~\cite{Ahrens:2011px,Cacciari:2008zb,Moch:2008qy, Kidonakis:2009mx,Kidonakis:2011jg} 
both at Tevatron~\cite{Aaltonen:2010bs,Abazov:2009si} and 
 at the LHC~\cite{ATLAS-CONF-2011-140,CMS-PAS-TOP-11-007}. Motivated by those results, the SM predictions for
 $t\bar t$ charge asymmetry have been revised in~\cite{Kuhn:2011ri,Hollik:2011ps} showing moderate $20\%$ increase due to QED and
 electroweak (EW) corrections.

 Numerous new physics (NP) scenarios (see~\cite{Westhoff:2011tq} for a recent  review) have been proposed to explain
the observed anomaly that predict the existence of new particles whose contributions induce the asymmetry.
Those scenarios can be directly tested at the LHC experiments by looking for new
 particle interactions. In the light of LHC results several popular explanations to the $t\bar t$ charge asymmetry such as
the axigluons~\cite{Ferrario:2008wm,Ferrario:2009bz,Rodrigo:2010gm}, $Z'$~\cite{Jung:2009jz} or $W'$~\cite{Cheung:2009ch} are stringently constrained.

However, effective field theory offers also model-independent tests of the top quark charge asymmetry.
Particularly interesting among those is the one due to the effective axial-vector coupling of the gluon~\cite{Gabrielli:2011jf}.
This scenario does predict the correct sign and the correct $m_{tt}$ dependence of the Tevatron anomaly and
does not necessarily require new light resonances. 
Therefore tests of this scenario  require particularly
precise measurement of the top quark charge asymmetry dependence on  $m_{tt}.$

As shown in ~\cite{Gabrielli:2011jf}, the characteristic 
new physics scale $\Lambda$, associated to a universal 
effective axial-vector coupling
of the gluon, should lie in a narrow range $\Lambda \simeq [1-1.3]$ TeV,
in order to correctly 
reproduce the Tevatron anomaly on top-quark charge asymmetry. The lower
bound on $\Lambda > 1$ TeV comes mainly by requiring conservative constraints 
on the total cross section of top-quark pair production, which is 
enhanced by the presence of an effective axial-vector coupling.
Therefore, the characteristic 
new physics scale associated to this scenario is within the discovery potential 
of LHC. We will discuss in more details this issue in section IV.

Although the LHC is $pp$ collider with symmetric initial state, the $t\bar t$ charge asymmetry can be
also defined and studied at the LHC in $q \bar q$ collisions using anti-quarks from the sea~\cite{Kuhn:1998jr,Kuhn:1998kw}. Because the sea quark
parton distributions differ from the valence ones, top and anti-top quark are preferably produced in different
rapidities. Therefore, as shown in~\cite{Kuhn:2011ri},
studying top quark charges asymmetries at large rapidities and large invariant masses
will enhance the asymmetries both in the SM as well as in NP scenarios.
At present the ATLAS~\cite{ATLAS-CONF-2011-106} and CMS~\cite{CMS-PAS-TOP-11-014} experiments have published only their measurements of rapidity cut-independent
 top quark charge asymmetries. Their results are consistent with the SM predictions. No rapidity nor invariant mass  dependent observables for top quark
 charge asymmetry  have been studied at the LHC, since the top-quark 
statistics is not yet large enough.

 The aim of this work is to study rapidity cut and invariant mass dependent top quark charge asymmetries at the LHC
 in the scenario of effective axial-vector coupling of the gluon. We show that the rapidity and invariant mass dependent
 asymmetries are much more appropriate for testing this type of NP scenarios than the cut-independent ones used by
the  ATLAS and CMS experiments so far. As a result,  the Tevatron observation of large $t\bar t$ charge asymmetry, 
that in $p\bar p$ collisions is equivalent to the $t \bar t$ forward-backward asymmetry $A_{FB}^t,$ 
can be either confirmed or ruled out already in the
 7~TeV LHC with 10~fb$^{-1}$ data.
In the latter case the LHC is able to put stringent constraint on the NP 
scale $\Lambda$ in this framework.

The paper is organized as follows. In section II we review the
theoretical framework of the effective field theory that generates
the axial-vector coupling of the gluon. In section III we study the effects of this scenario on the $t\bar t$ total cross section at the LHC.
In section IV we define different rapidity cut-dependent and independent charge asymmetries and present the results of our numerical studies.
We conclude in section V.

\section{Theoretical framework}
The most general effective Lagrangian for a generic quark-gluon
interaction, containing the lowest dimensional operators, and
compatible with gauge-, CP-, and Lorentz-invariance, is
\cite{Gabrielli:2011jf} \bea {\cal L}&=& -ig_S\Big\{ \bar Q
T^a\left[\gamma^\mu\left(1+ g_V(q^2,M)
+ \gamma_5 g_A(q^2,M)\right) G^a_\mu\right. \nonumber \\
&+&
\left.  g_P(q^2,M)q^{\mu}\gamma_5 G^a_\mu+
g_M(q^2,M) \sigma^{\mu\nu} G^{a}_{\mu\nu} \right] Q\Big\} ,
\label{vertex}
\eea
where $g_S$ is the strong coupling constant, $G^a_\mu$ and $G^a_{\mu \nu}$
are the gluon field and corresponding field strength respectively, 
$T^a$ are the color matrices,
$M$ is some energy scale, $q^2$ is the invariant momentum-squared carried by the gluon, and $Q$ denotes a generic quark field. Sum over the color index $a$ is 
understood.

At the moment we do not make any assumption on the origin of the form factors
$g_{A,P}(q^2,M)$ associated to the quark $Q$.
In the case of the SM, these
form factors are induced at 1-loop by the exchange of $W,Z$ bosons. In this
case the scale $M$ is connected to the EW scale, being related
to the $W,Z$ (masses) exchanged in the loop.
However, from now on we will assume that the dominant contribution
to the $g_{A,P}(q^2,M)$ arises from a NP which has a characteristic
scale above the EW scale. In this case the scale $M$ should be identified
with the NP scale.
The form factors $g_{A,P}$ depend also on  quark masses that
can be neglected for $m^2_Q\ll M^2$.
Finally,
the last term in Eq.~(\ref{vertex}) is the contribution of the chromomagnetic
dipole operator (with $g_M(q^2,M)$  the corresponding form factor), that may affect the total cross section~\cite{Haberl:1995ek,Hioki:2009hm} but 
does not significantly contribute to the asymmetry
$A_{FB}^t$ ~\cite{Blum:2011up}, and we shall not include it in our analysis. 

Model independently, the QCD gauge invariance implies a Ward identity (WI)  $2m_Q
g_A(q^2,M)=q^2 g_P(q^2,M)$, thus 
\bea \lim_{q^2\to 0}
g_{A,V}(q^2,M)=0\, ,\label{ginv} 
\eea 
since no $1/q^2$
singularities are present in $g_P$. 
Notice that the Ward Identity in Eq.(\ref{ginv}) is exact and free from 
any anomaly contribution, since the vector-axial coupling is an effective 
vertex and the fundamental theory (QCD) is anomaly free.
 As observed in
\cite{Gabrielli:2011jf}, Eq.~(\ref{ginv}) does not pose any
additional constraint on the form factors $g_{A,V}$, which could
have different magnitudes at arbitrary $q^ 2$. Therefore,
gauge-invariance does not prevent us to have $g_{V} \ll g_{A}$ as
long as $q^2 \neq 0$. We stress here once again that the QCD gauge
invariance is not broken and gluon remains massless because
$g_{A}$ and $g_{V}$ are induced via the form factors in
\Eq{vertex} that are subject to the condition in Eq.~(\ref{ginv}).

As stressed above, the  $g_{V,A}$ exist also in the SM, where they
are induced by EW radiative corrections, but are
numerically too small to have significant impact on the
observables we consider. However, if the origin of large
$A_{FB}^t$ observed at Tevatron is due to NP that has $(V\pm A)$ currents as in the SM,
large $g_{V}$ and $g_{A}$ can be generated. In
\cite{Gabrielli:2011jf} we found that this scenario is
phenomenologically unacceptable because $g_V$ is strongly
constrained by the total $q\bar q\to t\bar t$ cross section.
Indeed, being  $q\bar q\to t\bar t$ the dominant $t \bar t $
production mechanism at Tevatron, its cross section depends
quadratically on $g_A$ but only linearly on $g_V$. In particular,
the magnitude of $g_A$, necessary to explain the Tevatron
$A_{FB}^t$ anomaly, is not compatible with the condition $g_A \sim
g_V$, since $g_V$ is strongly constrained by the measurements on
the $p \bar p \to t \bar{t}$ cross section, which are in good
agreement with SM predictions. However, notice that the dominant
contribution to the $t \bar t $ production at LHC is given by the
$gg \to t \bar{t}$ process, which depends quadratically on $g_V$.
Therefore, the $t \bar{t}$ production cross section at LHC turns
out to be less sensitive to $g_V$ than at Tevatron.

Following the same approach as in \cite{Gabrielli:2011jf}, from
now on, we will neglect the contribution of the vectorial form
factor $g_V(q^2,M)$ in Eq.~(\ref{vertex}), and consider only NP
scenarios that generate $g_A$ with the hierarchy $g_V \ll g_A$. In
the limit of $q^2\ll M^2$, it is useful to parametrize the
axial-vector form factor as 
\bea 
g_A(q^2,M)=\frac{q^2}{\Lambda^2}
F(q^2,\Lambda)\, , 
\label{gA} \eea 
where  we absorb the NP
coupling $\alpha_{NP}$ and loop factor into the NP scale,
$\Lambda^2=M^2/(4\pi \alpha_{NP}).$ Because of the breaking of
conformal invariance, induced by renormalization, we expect
\cite{Raidal:1997hq} $F(q^2,\Lambda)$ to contain also logarithm
terms  $\log(q^2/\Lambda^2).$ This could give a large log
enhancement in the case of $|q^2|\ll \Lambda^2$. In general, the
form factor $F(q^2,\Lambda)$ could also develop an imaginary part
for $q^2>0$. In perturbation theory, this is related to the
absorptive part of the loop diagram generating $g_A$, when $|q^2|$
is above the threshold of some specific particles pair production.

\begin{figure}[t]
\begin{center}
\includegraphics[width=0.18\textwidth]{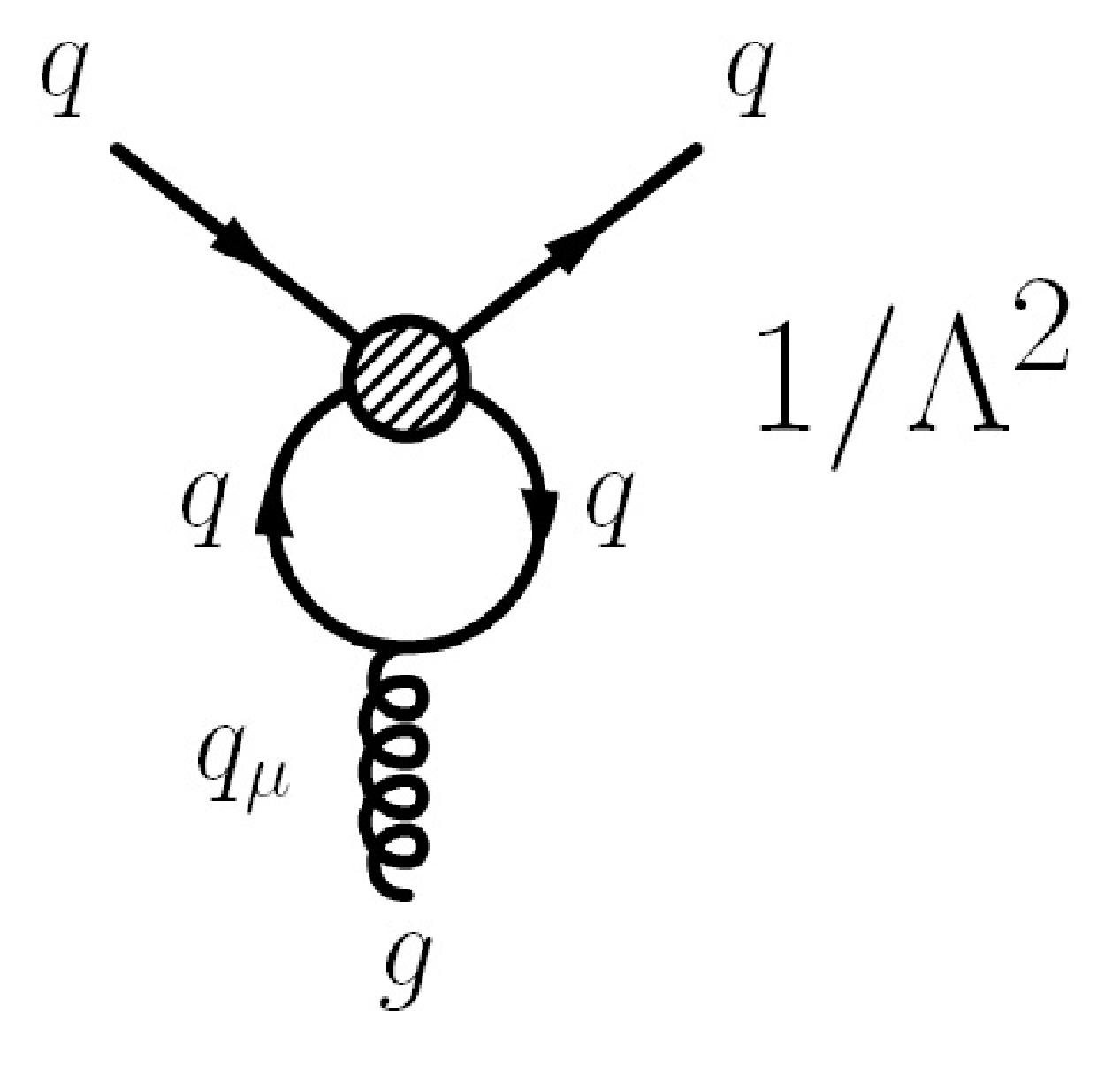}
\vspace{-0.3cm}
\caption{Feynman diagram in the effective low energy theory
that generates the effective axial-vector coupling of gluon, where 
$\Lambda$ is the scale related to the contact four-quarks operator.}
\label{fig:effvert}
\end{center}
\end{figure}

In \cite{Gabrielli:2011jf}, an origin of the anomalous large $g_A$
has been suggested. Assuming that there is a perturbative NP above
the EW scale, model independently the effective
operators~\cite{Delaunay:2011gv} \bea O^{1,8}_{AV}
&=&\frac{1}{{\Lambda}^2} [\bar Q T_{1,8}\gamma^\mu\gamma_5  Q]
[\bar Q T_{1,8} \gamma^\mu Q]\, ,
\label{OAV} \\
O^{1,8}_{PS}&=&\frac{1}{{\Lambda}^2}
[\bar Q T_{1,8}\gamma_5  Q] [\bar Q T_{1,8} Q] ,
\label{OPS}
\eea
 generate $g_A$ via 1-loop diagrams depicted in Fig.~\ref{fig:effvert}.
Here $T_1=1$ and $ T_8=T^a,$ thus both isoscalar and octet operators contribute. Notice that:
 $(i)$ no $g_V$ is induced due to the CP odd property of the operators in 
Eqs.(\ref{OAV}),(\ref{OPS}) and to QCD parity conservation;
$(ii)$ the 1-loop induced $g_A$ can be enhanced by large $\log(q^2/\Lambda^2)$,
although in this case the large logs need to be resummed 
via the renormalization group methods;
 $(iii)$ the operators $O^{1,8}_{AV}$, $O^{1,8}_{PS}$ do not induce FC processes; however, there could be different quark flavors in the loop
 in Fig.~\ref{fig:effvert} (extending the operator basis to $Q\to Q',$ $V\leftrightarrow A,$ $P\leftrightarrow S$ is straightforward);
 $(iv)$  the operators $O^{1,8}_{AV}$, $O^{1,8}_{PS}$ do not
interfere with the corresponding QCD induced 4-quark processes.
 The latter point has very important implications for our
scenario -- the stringent LHC
constraints~\cite{Khachatryan:2011as,Aad:2011aj} on
 4-quark contact interactions do not apply at all. Indeed, those constraints come from the interference between QCD and NP diagrams, and
 constrain the  models that  explain $A^t_{FB}$ with the similar interference very stringently. We stress that our scenario is free from those constraints
 and NP at 1-2~TeV can induce large $g_A$ as explained above.

However, the presence in the effective Lagrangian of the operators
in Eqs.~(\ref{OAV}),(\ref{OPS}) without
the counterparts of dimension-6 operators $O^{1,8}_{VV}$ and $O^{1,8}_{AA}$
(or analogously $O^{1,8}_{SS}$ and $O^{1,8}_{PP}$)
suppressed by a scale of the same order of
$\Lambda$, suggests more a scenario in which the above contact interactions
are a manifestation of quark compositeness,
rather than the exchange of a
heavy (perturbative) resonances among fundamental quark fields.
We do not elaborate on such a high scale NP model here.

Alternatively, large $g_A$ might be generated by new strongly-coupled
parity-violating dynamics related to EW symmetry breaking
(EWSB) at 1-2 TeV scale.
Because this NP is entirely non-perturbative,  generating $g_A$ is possible~\cite{Lane:1996gr}
but we are not able to compute it. We are only able to estimate
the validity range of the effective coupling parametrization
in Eq.~(\ref{gA}) that is controlled by
$\hat{s}/\Lambda^2_{eff}$, where $\Lambda^2_{eff}$  is
expected to be related to $\Lambda$ as
$\Lambda_{eff} \sim \Lambda/\sqrt{\alpha_S}$.
For $\Lambda \sim 1 (1.3)$ TeV, as required by the
$A^t_{FB}$ anomaly, the related scale $\Lambda_{eff}$ is 3.5 (4.6) TeV.
At this scale a plethora of new resonances should occur
at the LHC allowing to test this scenario.
Notice that, in the region of large invariant masses
$\hat{s}\gg \Lambda_{eff}^2$, the low-energy ansatz
$g_A\sim q^2/\Lambda^2$ is not valid anymore and the $q^2$
dependence of $g_A$ should be determined by fitting the data.
Unitarity should require that $g_A\le 1 $ at large exchanged
momenta $|q^2|\gg \Lambda^2$, bounding the anomalous
behavior of the total cross section with energy.

\section{Cross sections}
Here we analyze the contribution of the axial-vector
$g_A$ anomalous coupling,
as defined in Eq.~(\ref{vertex}), to the partonic cross sections
for $t\bar t$ pair production at the LHC,
related to the processes  $q\bar{q}\to t\bar t$ and $g g\to t\bar t$.

\subsection{$q\bar{q}\to t\bar t$ process}
\begin{figure}[t]
\begin{center}
\includegraphics[width=0.45\textwidth]{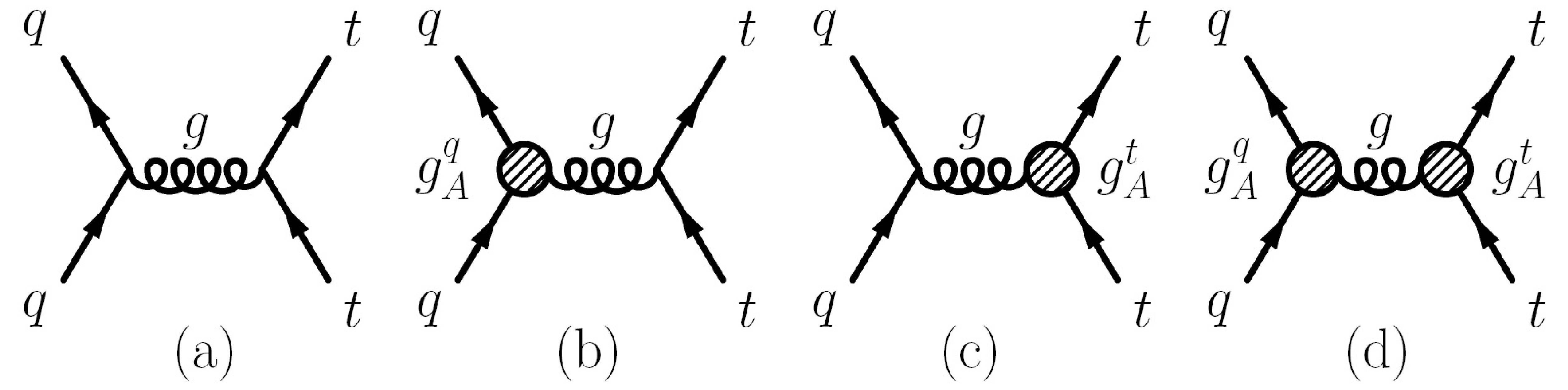}
\vspace{-0.3cm}
\caption{Feynman diagrams (a)-(d)
for the $q \bar q \to t \bar t$ process, with
the contribution of the gluon effective axial-vector couplings $g^{q,t}_A$.}
\label{fig:qqtt}
\end{center}
\end{figure}

Let us consider the tree-level scattering
\bea
q(p_1) \bar{q}(p_2)\to t(p_3)\bar t(p_4)\, ,
\eea
where $p_{1-4}$ are the corresponding particles momenta and $q$ stands for
a light quark.
The Feynman diagrams (a)-(d) relative to
$q\bar q \to t\bar t$, including the axial-vector coupling,
are shown in Fig. \ref{fig:qqtt}. According to Eq.~(\ref{vertex}), supplemented
by the Ward identity in Eq.~(\ref{ginv}),
the Feynman rule $\Gamma_A^{a~ \mu}$,
corresponding to the effective axial-vector gluon couplings to quarks $q$ is
\bea
\Gamma_A^{a~ \mu}=
i g^q_A\,
T^a\left(\gamma_{\mu}\gamma_5 -2 q_{\mu}\frac{m_q}{q^2}\gamma_5\right)\, ,
\label{Frule}
\eea
where $q_{\mu}$ is the gluon momentum entering the vertex, $m_q$ is the
quark mass, and $T^a$ the color matrix. From now on, we will omit the $q^2$
dependence in the $g^q_A$ form factors, unless specified.
The corresponding differential cross section in the massless
light-quarks $q$ limit (summed over all colors) is given by
\bea
\!\!\!
\frac{d \sigma^{q\bar q}}{d \hat t}&=&
\frac{8 \pi \alpha_S^2}{9\hat{s}^4}
\Big[\Big(\hat{t}^2 +(\hat{s}-2 m_t^2)\hat{t}+\frac{\hat{s}^2}{2}
+m_t^4\Big)\times
\nonumber\\
&&
\left(
1+|g_A^q|^2+|g_A^t|^2 + |g_A^q|^2|g_A^t|^2 \right)
\nonumber \\
&-&2 |g_A^t|^2 \,m_t^2\hat{s} \left(1+|g_A^q|^2\right)
\nonumber \\
&+& 2 {\rm Re}[g_A^q]\, {\rm Re}[g_A^t]\, \hat{s}(\hat{s}+2\hat{t}-2m_t^2)
\Big]\, ,
\label{dsigmaqq}
\eea
where $g_A^q$ and $g_A^t$ are the corresponding axial-vector form factors
for the light-quark $q$ and top-quark, respectively.
The Mandelstam variables $\hat  s,\hat t,\hat u$ are defined as
\bea
\hat s=(p_1+p_2)^2, ~~~ \hat t=(p_1-p_3)^2, ~~~ \hat u=(p_1-p_4)^2\, .
\label{mandelstam}
\eea
The result in Eq.~(\ref{dsigmaqq}) is gauge invariant due to the
corresponding Ward identity in Eq.~(\ref{ginv}).
After integrating Eq.~(\ref{dsigmaqq}) over the full range of $\hat{t}$,
the total partonic cross section is given by \cite{Gabrielli:2011jf}
\bea
\sigma^{q\bar q}(\hat s)&=&
\frac{8\pi\alpha^2_S \beta_t}{27 \hat{s}}\Big\{(1+2\frac{m_t^2}{\hat s})
\left(1+|g_A^q|^2\right)+
\nonumber \\
&&
\beta_t^2 |g_A^t|^2\left(1+|g_A^q|^2\right)\Big\} ,
\label{sigmaqq}
\eea
where $\beta=\sqrt{1-\rho}$ and $\rho=4m_t^2/\hat{s}$.

\subsection{$gg\to t\bar t$ process}
\begin{figure}[t]
\begin{center}
\includegraphics[width=0.45\textwidth]{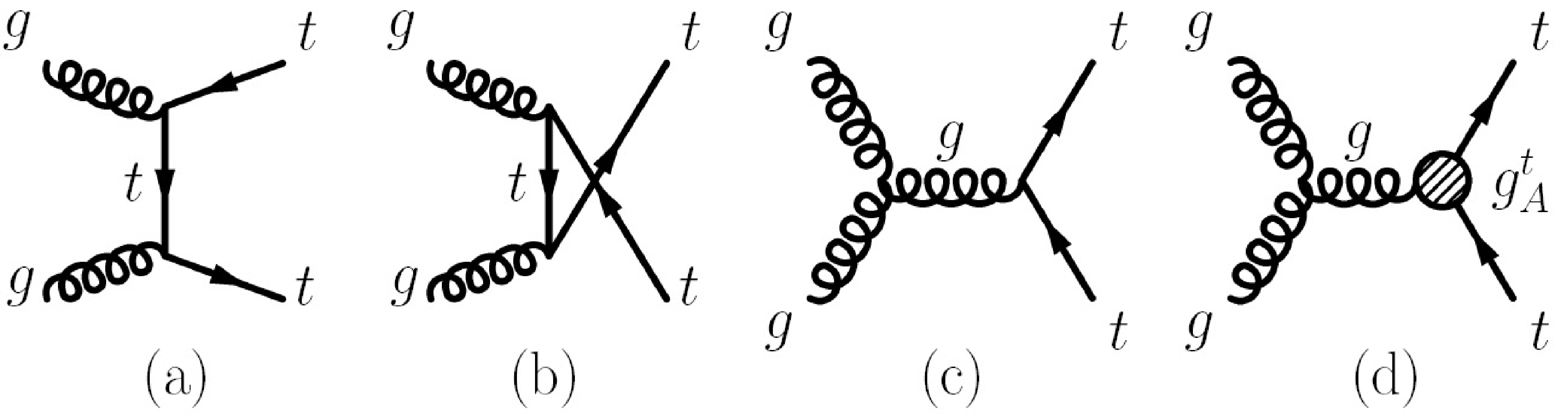}
\vspace{-0.3cm}
\caption{Feynman diagrams (a)-(d)
for the $g g \to t \bar t$ process, with
the contribution of the gluon effective axial-vector coupling $g^{t}_A$.}
\label{fig:ggtt}
\end{center}
\end{figure}

The Feynman diagrams relative to the tree-level process
$gg\to t\bar t$ are shown in Fig.~\ref{fig:ggtt}.
Here, the effective axial-vector coupling of the top-quark
affects only the s-channel (diagram \ref{fig:ggtt}(d) ).
Indeed, due to the Ward identity in Eq.~(\ref{ginv}),
the effective axial-vector contribution
vanishes in the $\hat t$- and $\hat u$-channels,
diagrams \ref{fig:ggtt}(a) and \ref{fig:ggtt}(b) respectively,  because the gluons
attached to the axial-vector vertex are on-shell.

Finally, the differential unpolarized cross section (summed over all colors)
for the $g g\to t \bar{t}$ scattering,
corresponding to the Feynman diagrams in Fig.~\ref{fig:ggtt}, is given by
\bea
\frac{d \sigma^{gg}}{d\hat{t}}&=&
\frac{\pi \alpha_S^2}{64 \hat{s}^2} \left[
12 M_{ss}+\frac{16}{3} \left(M_{tt}+M_{uu}\right)
-\frac{2}{3} M_{tu} \right. \nonumber \\
&+& \left. 6\left(M_{st}+M_{su}\right)\right],
\label{dsigmagg}
\eea
with
\bea
M_{ss}&=&\frac{4}{\hat{s}^2}\left[
\left(\hat{t}-m_t^2\right)\left(\hat{u}-m_t^2\right)
+\left(\hat{t}\hat{u}-m_t^4\right)|g_A^t|^2\right],
\nonumber \\
M_{tt}&=&\frac{2}{(\hat{t}-m_t^2)^2}
\left[\left(\hat{t}-m_t^2\right)\left(\hat{u}-m_t^2\right)
-2m_t^2 \left(\hat{u}+m_t^2\right)\right],
\nonumber \\
M_{tu}&=&\frac{4m_t^2}{(\hat{t}-m_t^2)(\hat{u}-m_t^2)}
\left(\hat{s}-4m_t^2\right),
\nonumber \\
M_{st}&=&\frac{4}{\hat{s}(\hat{t}-m_t^2)}\left[m_t^4-
\hat{t}\left(\hat{s}+\hat{t}\right)\right]\, ,
\label{M2gg}
\eea
and $M_{uu}=M_{tt}\left\{ t\leftrightarrow u \right\}$,
$M_{su}=M_{st}\left\{ t\leftrightarrow u \right\}$.

Although the effective axial-vector coupling affects only the
$M_{ss}$ contribution proportional to the $|g^t_A|^2$ term in Eq.~(\ref{M2gg}),
this contribution is actually $SU(3)_c$ gauge invariant.
In order to understand that, let us decompose the
amplitude for the $gg\to t \bar{t}$ process as
\bea
M=M^{QCD}+M^A_{s}\, ,
\label{AMPL}
\eea
where $M^{QCD}$ represents the full QCD contribution to the total amplitude
and $M^A_s$ is the diagram contribution
in s-channel diagram \ref{fig:ggtt}(d) proportional to
the effective axial-vector coupling $g_A$, obtained by using
the standard QCD Feynman rules for the 3-gluon vertex.
While it does not depend on the $SU(3)_c$ gauge-fixing,
due to the conservation of the
effective axial-vector vertex,
the $M^A_s$ diagram alone is not manifestly $SU(3)_c$ gauge invariant under the
gauge transformations on external states, namely
$\epsilon_{\mu}^a(p_1)\to p_{1\mu}$
and $\epsilon_{\mu}^b(p_2)\to p_{2\mu}$, where $\epsilon_{\mu}^a(p_1)$
and $\epsilon_{\mu}^b(p_2)$ indicate the polarization vectors of initial 
gluons (here $a,b$ stand for color indices). This non-invariance is
due to the presence of the QCD 3-gluon
vertex in the diagram \ref{fig:ggtt}(d).
The gauge-dependent part of the  $M^A_s$ amplitude (proportional to $g_A$)
is not canceled by the corresponding
$\hat t$- and $\hat u$-channels, since the  $g_A$ contribution to these
channels is vanishing due to the condition in Eq.~(\ref{ginv})
for on-shell external gluons.
However, this is not a problem and it is an artifact of the effective
theory. Indeed, one can always construct a full amplitude for the
$gg\to q\bar q$ process including the $g_A$ contribution
in a manifestly gauge invariant way.
Now we will prove that the result of Eq.~(\ref{dsigmagg}), obtained by
using the $M$ amplitude in Eq.~(\ref{AMPL}), can also be obtained by using
a manifestly gauge invariant amplitude $M_{GI}$.

In order to show that, let us add to $M$ a new contribution
$\bar{M}^A_s$,  which is identical to the s-channel diagram $M^A_s$ contribution
in Fig. \ref{fig:ggtt}(d), but with the 3-gluon vertex suitable
modified.
In particular, in $M^A_s$  the Lorentz structure of the 
QCD 3-gluon vertex (in momentum space)
$\Gamma_{\scriptscriptstyle{\rm QCD}}^{\alpha \beta \mu}$ will be replaced by
a new 3-gluon vertex
$\bar{\Gamma}^{\alpha\beta \mu}$ defined as
\bea
p_{1\alpha}\left(\Gamma_{\scriptscriptstyle{\rm QCD}}^{\alpha\beta \mu}+\bar{\Gamma}^{\alpha\beta\mu}
\right)=
p_{2\beta}\left(\Gamma_{\scriptscriptstyle{\rm QCD}}^{\alpha\beta \mu}+\bar{\Gamma}^{\alpha\beta\mu}
\right)=0\, + \cdots,
\label{WI}
\eea
where $\cdots$ stands for terms proportional to $p_{2\beta}$
and/or $p_{1\alpha}$, that vanish when contracted with the
external on-shell gluon polarizations
$\epsilon^a_{ \alpha}(p_1)$ and
$\epsilon^b_{ \beta}(p_2)$ respectively.
It is easy to show that the required expression for
$\bar{\Gamma}^{\alpha\beta \mu}$ is given by
\bea
\bar{\Gamma}^{\alpha\beta \mu}=\frac{2}{s}p_2^{\alpha}p_1^{\beta}
\left(p_2^{\mu}-p_1^{\mu}\right)+
2\delta^{\mu\alpha}p_1^{\beta}-2\delta^{\mu\beta}p_2^{\alpha}\, ,
\label{3G}
\eea
where, in Eq.~(\ref{3G}), the indices $\alpha$ and $\beta$ are understood to
be contracted with the on-shell gluon polarization vectors
$\epsilon_{\alpha}^a(p_1)$ and $\epsilon_{\beta}^b(p_2)$ respectively,
with $p_1$ and $p_2$ momenta entering the 3-gluon vertex.
When this new diagram $\bar{M}_s^A$ is added to the total amplitude $M$
in Eq.~(\ref{AMPL}) the effective amplitude
$M_{GI}= M+\bar{M}_s^A$ turns out to be manifestly gauge invariant
under the transformations $\epsilon_{\mu}^a(p_1)\to p_{1\mu}$
and $\epsilon_{\mu}^b(p_2)\to p_{2\mu}$ due to the relation (\ref{WI}).
Finally, after a bit of algebra, one can prove that the following
relation holds
\bea
\sum_{\rm pol}|M|^2=\sum_{\rm pol}|M_{GI}|^2\, ,
\label{MR}
\eea
showing that the result for the cross-section
in Eqs.~(\ref{dsigmagg}),(\ref{M2gg}),
including the term in $M_{ss}$ proportional to $|g_A|^2$, is
truly $SU(3)_c$ gauge invariant.

After integrating Eq.~(\ref{dsigmagg}) over the full range of $\hat{t}$,
we obtain for the total partonic cross section
\bea
\hat{\sigma}^{gg}(\hat{s})&=&\frac{\pi\alpha_S^2}{48\hat{s}}
\Big\{\left(16+\rho\left(16+\rho\right)\right)
\log{\left(\frac{1+\beta}{1-\beta}\right)}
 \nonumber\\
&-& \beta\left(28+31\rho+6|g_A|^2\left(\rho-1\right)\right)\Big\}\, .
\label{sigmagg}
\eea

Our results for the cross sections 
appearing in Eqs.(\ref{dsigmaqq}), (\ref{sigmaqq}), 
and (\ref{dsigmagg}), (\ref{M2gg})
are consistent with the corresponding QCD results 
\cite{Beenakker:1993yr} in the limit of $g_A^{q,t}\to 0$. Moreover, 
Eqs.(\ref{dsigmaqq}), (\ref{sigmaqq}) are consistent 
with the corresponding results in the axigluon models \cite{Ferrario:2008wm}, 
in the limit of vanishing axigluon mass and for the part concerning the
vector-axial couplings.

Finally, the hadronic cross section  $pp\to t \bar t X$ at LHC is
obtained by convoluting the partonic cross sections in Eqs.
(\ref{sigmaqq}),(\ref{sigmagg}) with the corresponding parton
distribution functions (PDF) for quarks and gluons, namely \bea
\sigma_{p p \to t\bar t X}={\int \left( \sum_qd \mu_q
\sigma_{qq}(\hat s)+ d\mu_{g}\sigma_{gg}(\hat s)\right)},
\label{xsec} \eea where $d \mu_{q}$ and $d \mu_{g}$ indicate the
differential integrations in $dx_1 dx_2$ convoluted with the
quarks and gluon PDF, respectively. In the numerical integration of
Eq.~(\ref{xsec}) we have used the CTEQ6L1 parton distribution
function (PDF) \cite{Pumplin:2002vw}, where we set the PDF scale
$\mu=m_t$ with top-quark mass $m_t=172$ GeV.

\section{Charge asymmetries at the LHC}

\subsection{Definitions}
Here we present the numerical results
for the gluon axial-vector contribution to the $t \bar t$ charge asymmetry
at the LHC. Let us first review the SM contribution to charge asymmetry.

In the SM, the angular and rapidity distributions of the top and
anti-top quarks are identical at tree-level. However, a
$t \bar t$ charge asymmetry (of order ${\cal O}(\alpha_S^3)$
can be generated at one-loop level by the interference of the tree-level
diagram for the $q\bar q\to t \bar t$ process with the corresponding
one-loop QCD box-contribution \cite{Kuhn:2011ri}.
In particular, QCD predicts that
top-quarks become more
abundant in the direction of the incoming light quarks.
The EW interactions, with
the Z-boson exchange in the s-channel, does not contribute at the tree-level
since the interference with the QCD s-channel diagram is vanishing because
of the singlet color representation of the Z boson.
At the Tevatron, the top-quark charge asymmetry is equivalent to
the forward-backward (FB) asymmetry due to the charge conjugation symmetry
of the initial $p \bar p$ state.

On the other hand, due to the symmetry of the colliding initial
proton-proton state,
the top-quark production at LHC is forward-backward symmetric in the
laboratory frame. This means that, when integrated over the full kinematic
range, the top-quark charge asymmetry vanish, as well as the the FB asymmetry.
Nevertheless, it is still possible to get a non-vanishing
charge asymmetry in suitable kinematic regions.
The physical reason can be understood as follows.
According to QCD, top-quarks are
preferentially emitted in the direction of the incoming quarks.
Since quarks have larger momenta than anti-quarks in the proton, the
asymmetry at partonic level is transformed into an excess of top quarks in the
forward and backwards regions due to the boost into the laboratory frame.
This suggests that, when we restrict the sample of events to specific kinematic
regions, a non-vanishing charge asymmetry can measured at the LHC.

Following the conventions adopted in Ref.~\cite{Kuhn:2011ri}, we
are considering here the following set of charge asymmetries that
can be measured at the LHC, defined as
\begin{itemize}
\item the {\bf in} and {\bf out} {\it cut-dependent}  charge asymmetries
\bea
A_C^{\rm in}(y_C)= \frac{N(|y_{\bar{t}}|< y_C)-N(|y_t|< y_C)}{
 N(|y_{\bar{t}}|< y_C)+N(|y_t|< y_C)}\, ,
\label{CAin}
\eea
\bea
A_C^{\rm out}(y_C)= \frac{N(|y_{\bar{t}}|> y_C)-N(|y_t|> y_C)}{
 N(|y_{\bar{t}}|> y_C)+N(|y_t|> y_C)}\, ,
\label{CAout}
\eea
as a function of the cut $y_c$ on the top $y_t$ and anti-top $y_{\bar t}$
quarks rapidities;
\item
the {\it cut-independent} charge asymmetry, as measured by ATLAS and CMS,
\bea
A_C= \frac{N(\Delta_y > 0)-N(\Delta_y < 0)}{N(\Delta_y > 0)+N(\Delta_y < 0)},
\label{CA}
\eea
where $\Delta_y \equiv |y_t|-|y_{\bar t}|$;
\item the {\it cut-dependent pair} charge-asymmetry
\bea
A^{\rm cut}_C(Y_c)= \frac{N(y_t > y_{\bar t})-N(y_t < y_{\bar t} )}
{N(y_t > y_{\bar t})+N(y_t < y_{\bar t} )},
\label{CAcut}
\eea
as a function of the cut $Y_c$ on mean rapidity, namely
$(y_t+y_{\bar t})/2 > Y_c$.
\end{itemize}
All the above observables are defined in the laboratory frame.
Due to the symmetry of the initial proton-proton configuration,
both $A_C^{\rm in/out}(y_C)$ and $A_C^{\rm cut}(Y_c)$
vanish if the whole rapidity spectrum is integrated, that is when
$y_c$ and $Y_c$ approach their maximum kinematic allowed values.

The top-quark production by the gluon-gluon fusion mechanism,
which is dominant at the LHC (namely $70\%$
and $90\%$ at 7 TeV and 14 TeV c.o.m. energy respectively), is charge symmetric
under higher order corrections. Moreover, as can be seen from
Eqs.~(\ref{dsigmagg})-(\ref{M2gg}),
this mechanism remains charge symmetric also in the presence of an effective
axial-vector coupling contribution. Therefore, the charge antisymmetric
contributions to top quark production are thus screened
at the LHC mainly due to the gluon-gluon fusion. However, the
contributions of the gluon-gluon collisions can be
reduced by imposing a lower cut on the top-pair invariant mass
$m_{tt}$. This has the effect of eliminating
the regions of  lower longitudinal momentum fraction of the colliding partons
where the gluon density is much larger that the quark densities.

By imposing lower cuts on $m_{tt}$
has also the advantage of enhancing the $q\bar q \to t \bar t $
contribution to the charge asymmetry, although at the price of
reducing the statistics of $t \bar t$ pairs.
As we will show
in the following, this requirement has also crucial implications
for our scenario, since it increases
the contribution of the axial-vector coupling of gluon to the charge
asymmetry. This is due to the fact that the effective coupling
$g_A$ grows as $m^2_{tt}/\Lambda^2$ at large $m_{tt}$ values, but still
$m_{tt}\le \Lambda$.

Following the results of \cite{Gabrielli:2011jf}, we will restrict
our analysis to the case of real and universal axial-vector gluon
couplings. In particular, by neglecting higher order terms in $q^2$,
we parametrize at low energy the axial-vector coupling
(for values of $|q^2|< \Lambda^2 $ ) as follows 
\bea
g_A^t=g_A^q=\frac{q^2}{\Lambda^2}\, \, , 
\label{univ}
\eea 
which corresponds to neglect the $q^2$ dependence in the form factor $F(q^2,M)$
in Eq.(\ref{gA}) and normalize it to 1. 
All NP couplings are then absorbed 
in the scale $\Lambda$.

Indeed, in order to explain the Tevatron anomaly on top-quark FB
asymmetry, while requiring conservative constraints on the $t \bar
t$ cross sections at Tevatron, in \cite{Gabrielli:2011jf} it was
suggested that the most favored scenario is the one where all
axial-vector couplings are universal and real, with the NP scale
$\Lambda$ that lies in a narrow range $1~ {\rm TeV} < \Lambda <
1.3~ {\rm TeV}$.

However, we stress that the role of new free parameters proportional to
the imaginary part of the axial-vector form factors,
should not dramatically affect our results.
The reason is the following. The charge asymmetry is directly proportional to
the real part of the $g_A$ form factors, in particular to
Re[$g_A^q$]Re[$g_A^q$], cfr. the last
term in the right hand side (r.h.s.) of Eq.~(\ref{dsigmaqq}).
On the other hand, the Im[$g_A$] enters only through $|g_A|$ in
the denominators of Eqs.~(\ref{CAin})-(\ref{CAcut}),
thus it affects only the total cross section or
analogously the total number of events in the asymmetry.
Moreover, the NP contribution to the total $t\bar t $ cross section
at LHC is largely
screened by the QCD gluon-gluon production mechanism.
Therefore, by requiring conservative constraints on the total cross section,
the dependence of the charge-asymmetry by Im[$g_A$] will be strongly limited,
justifying in part the fact that Im[$g_A$] does not play a crucial
role in the present analysis.

In our numerical analysis we have not included the SM contribution to the
charge asymmetry. Indeed, this is almost negligible with respect
to the axial-vector gluon contribution for most of the
kinematic regions considered here. In particular, we have retained only
the $g_A$ contribution in the numerators of the r.h.s.
of Eqs.~(\ref{CAin})-(\ref{CAcut}), neglecting the corresponding
SM contribution. Clearly, we have retained the SM effect, at the leading 
order (LO) in QCD, in the evaluation of the total number of events entering in
the equation for the charge asymmetry.

\begin{figure*}[t]
\begin{center}
\includegraphics[width=0.40\textwidth, angle=-90]{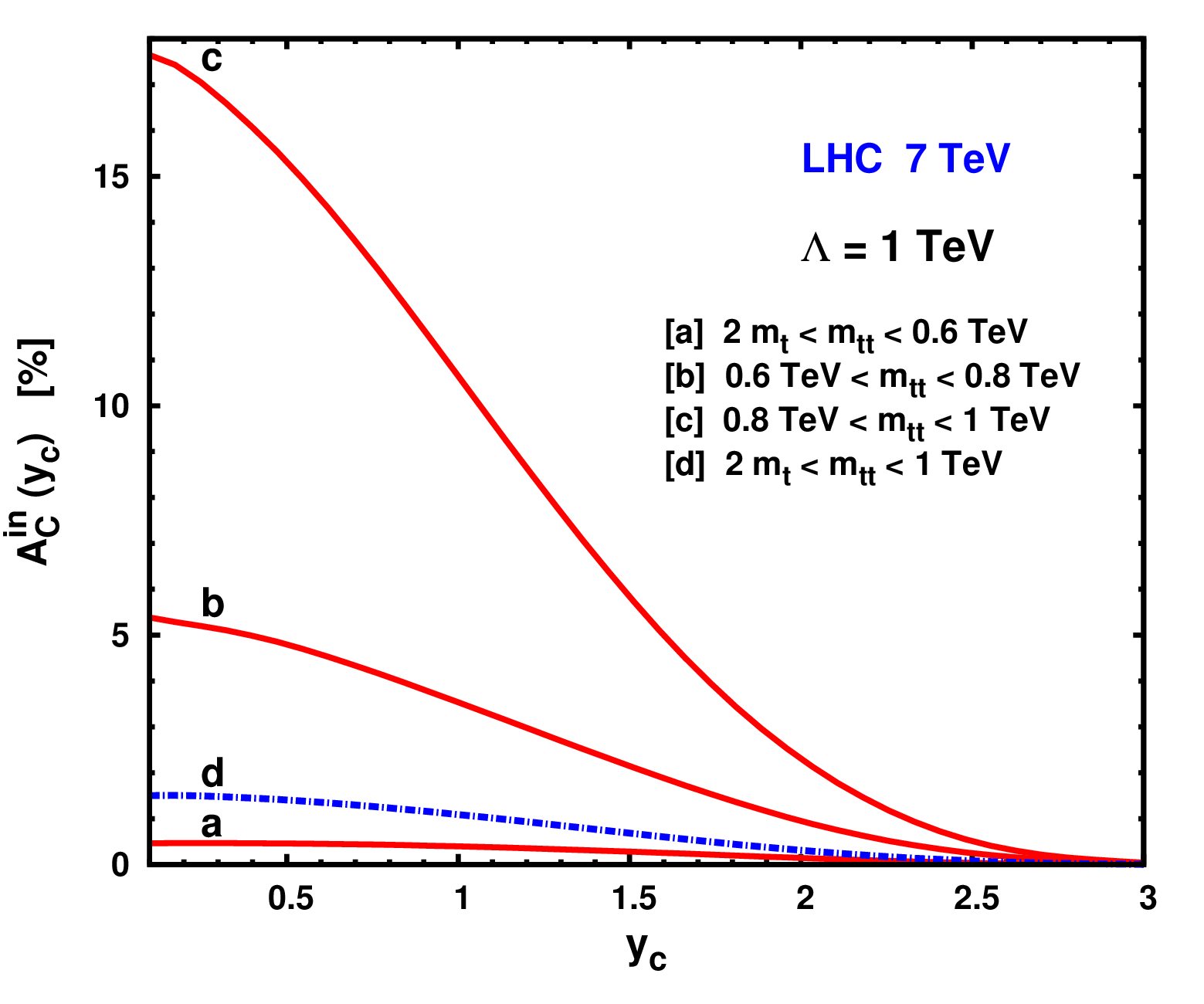}
\includegraphics[width=0.40\textwidth, angle=-90]{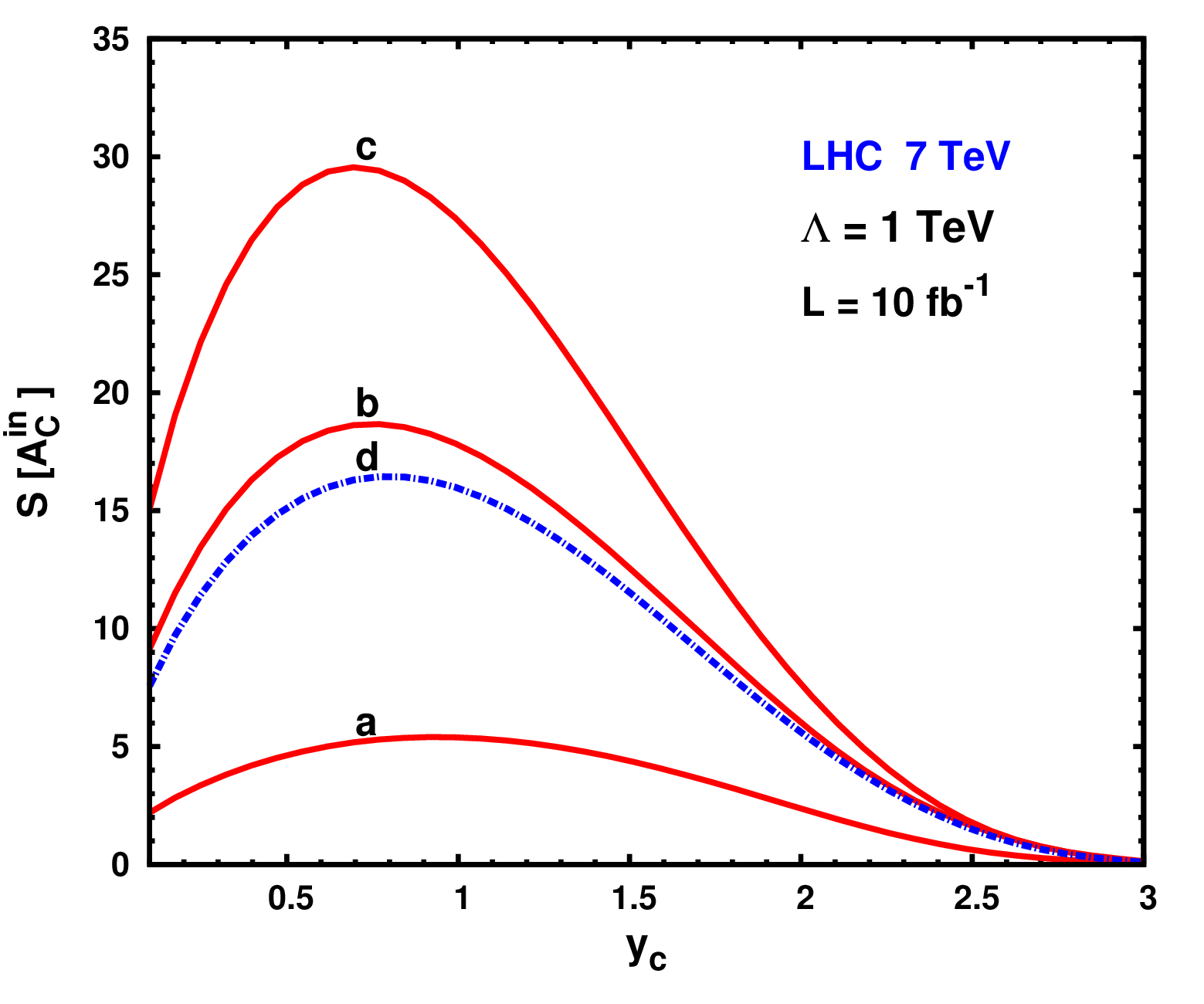}
\includegraphics[width=0.40\textwidth, angle=-90]{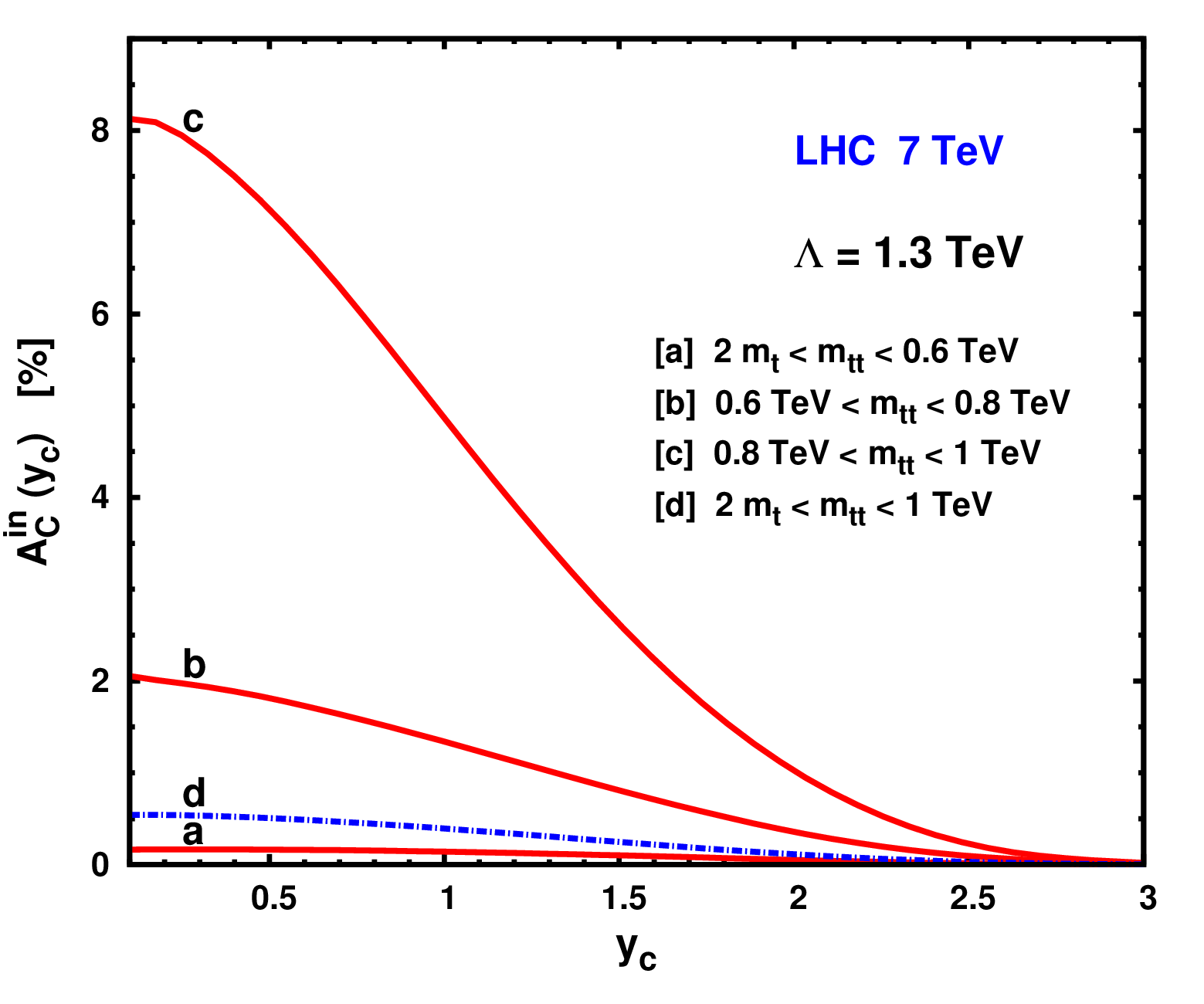}
\includegraphics[width=0.40\textwidth, angle=-90]{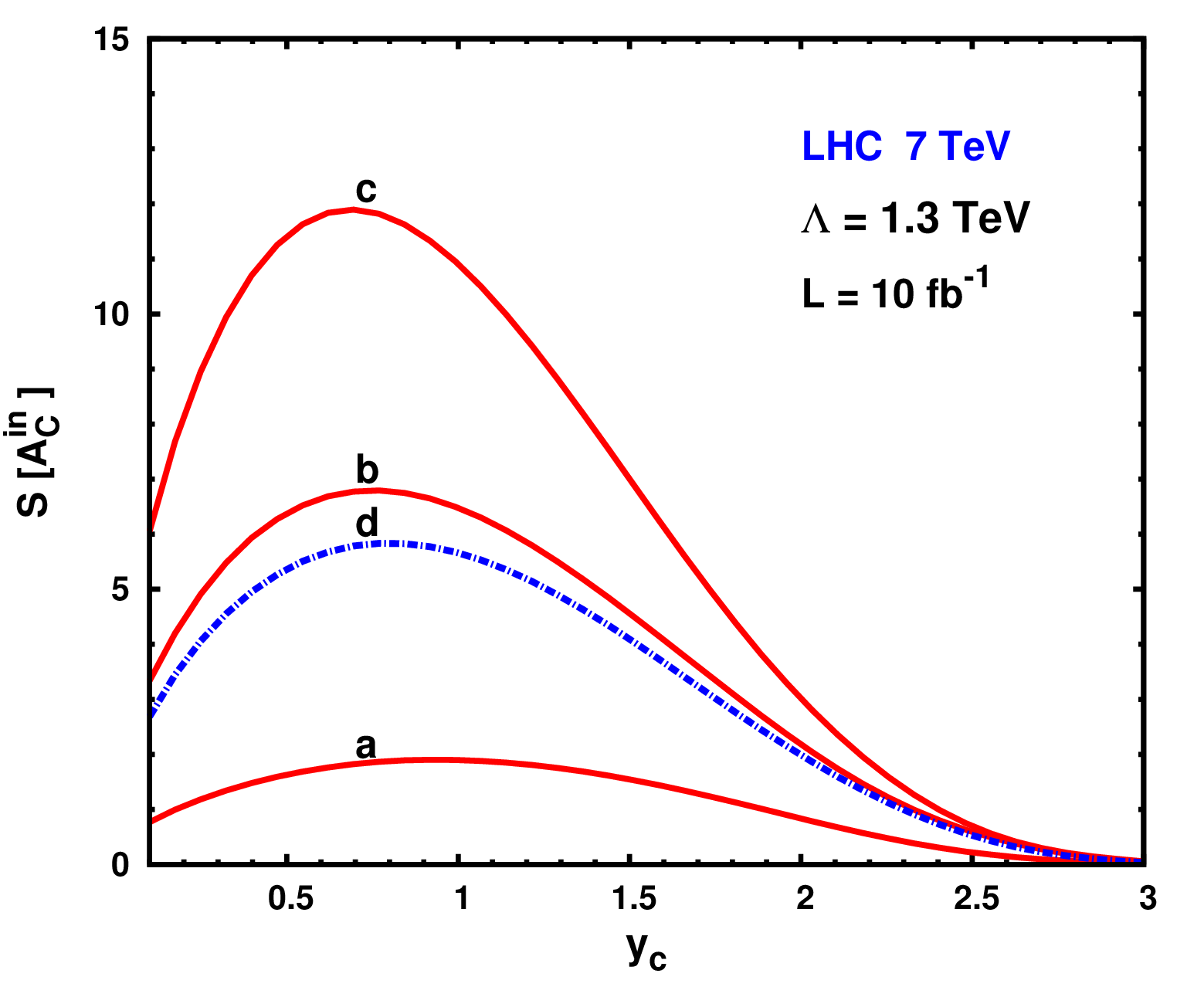}
\vspace{-0.3cm}
\caption{The $t\bar t$ charge asymmetry $A_{C}^{in}(y_c)$
in percentage  (left plots)
and corresponding statistical significance $S[A_C^{in}]$ (right plots)
at LHC with $pp$ center of mass energy $\sqrt{S}$ = 7 TeV
and integrated luminosity $L=10~{\rm fb}^{-1}$, with $m_t=172$ GeV,
as a function of the cuts on the $t$- and $\bar t $-quark rapidity $y_c$
(in the lab frame) and
for several regions ([a-d]) of $t\bar t$ invariant mass $m_{tt}$.
Up and down plots correspond to the scale $\Lambda=1$ TeV
and $\Lambda=1.3$ TeV,  respectively.}
\label{fig:ACin}
\end{center}
\end{figure*}

Following the definition of asymmetry,
the full value of $A_C^{\rm SM+ NP}$, including the SM one
($A^{\rm SM}_C$), can be related to the results of $A_C^{NP}$ presented
here by the following relation
\bea
A_C^{\rm SM+ NP}=A^{NP}_C+\frac{A_C^{\rm SM}}{1+\Delta \sigma}\, ,
\label{asym}
\eea
where, $\Delta \sigma$ is
the percentage variation of the total cross section (defined as
$\Delta \sigma=\sigma^{\rm NP}/\sigma^{SM}$), and $\sigma^{\rm NP~ (SM)}$ 
represent the pure NP (SM) contributions to the total cross
sections evaluated in the same kinematic region of
charge asymmetry. The numerical values of $\Delta \sigma$,
as function of $\Lambda$ and for some kinematic regions of $m_{tt}$,
are plotted in Fig.~\ref{fig:deltasigma}. The symbol $A_C$ appearing in the
figures stands for the pure NP contribution to the charge
asymmetry $A^{\rm NP}_C$ as defined above.

\subsection{Numerical results}
\begin{figure*}[t]
\begin{center}
\includegraphics[width=0.40\textwidth, angle=-90]{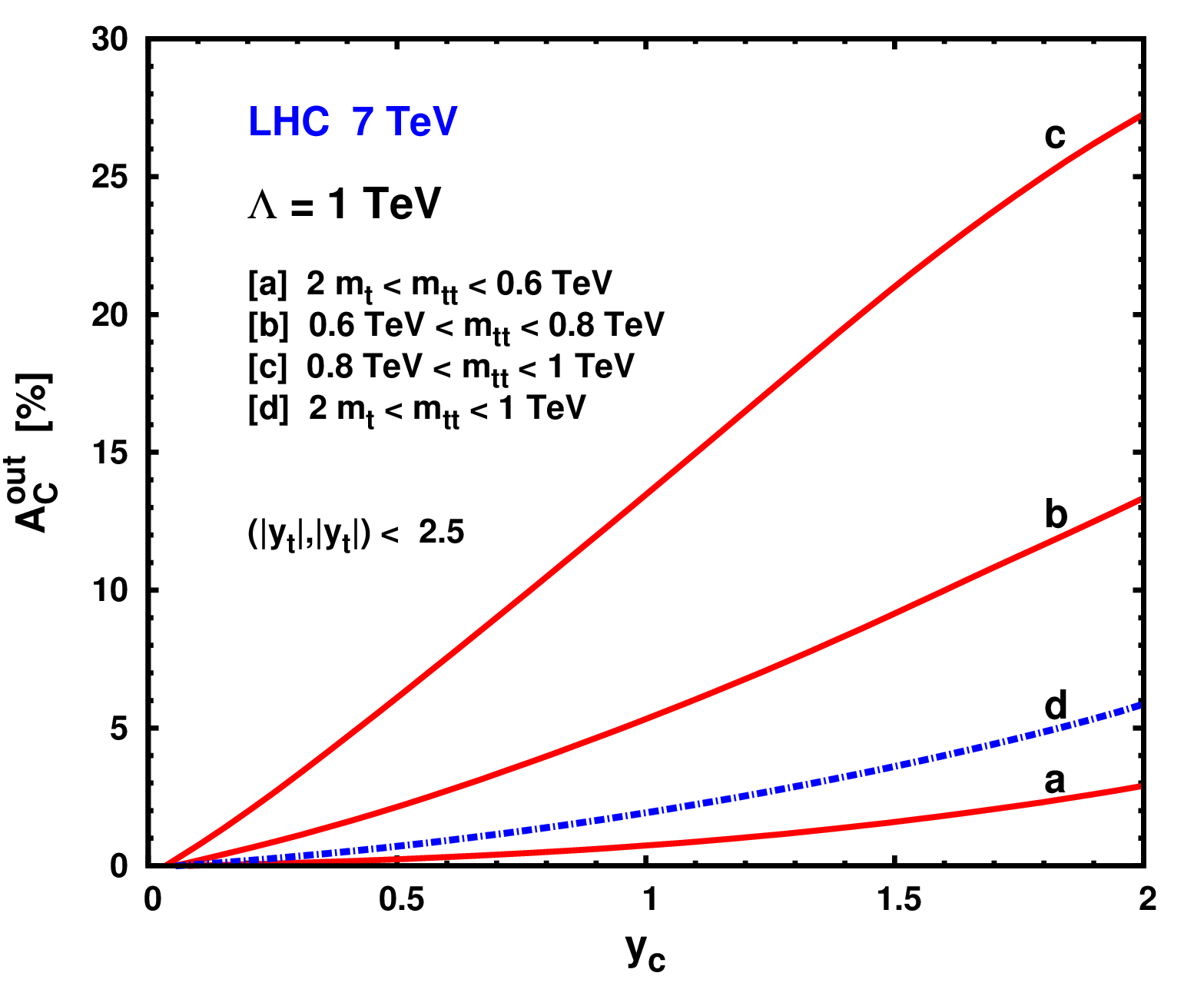}
\includegraphics[width=0.40\textwidth, angle=-90]{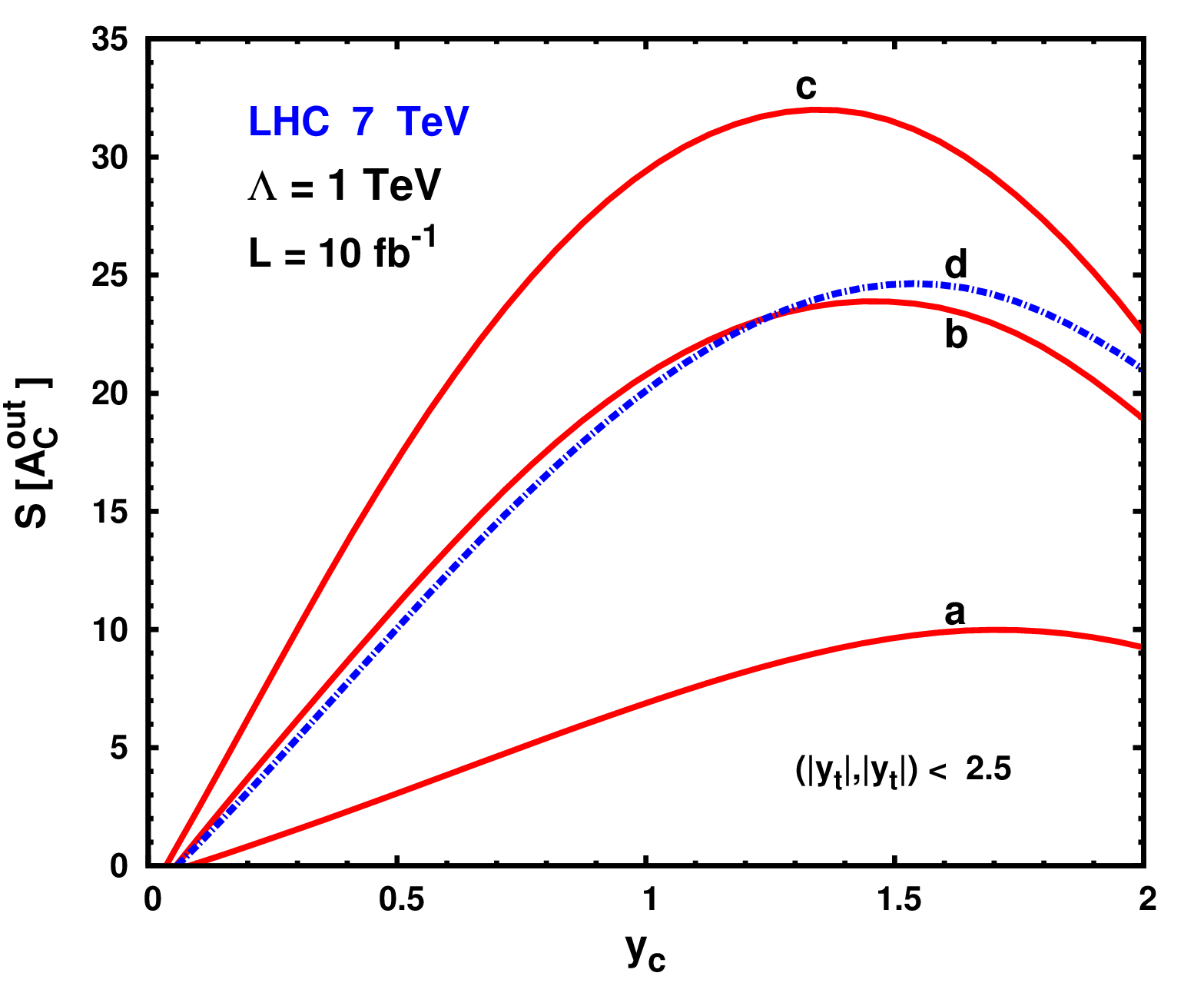}
\includegraphics[width=0.40\textwidth, angle=-90]{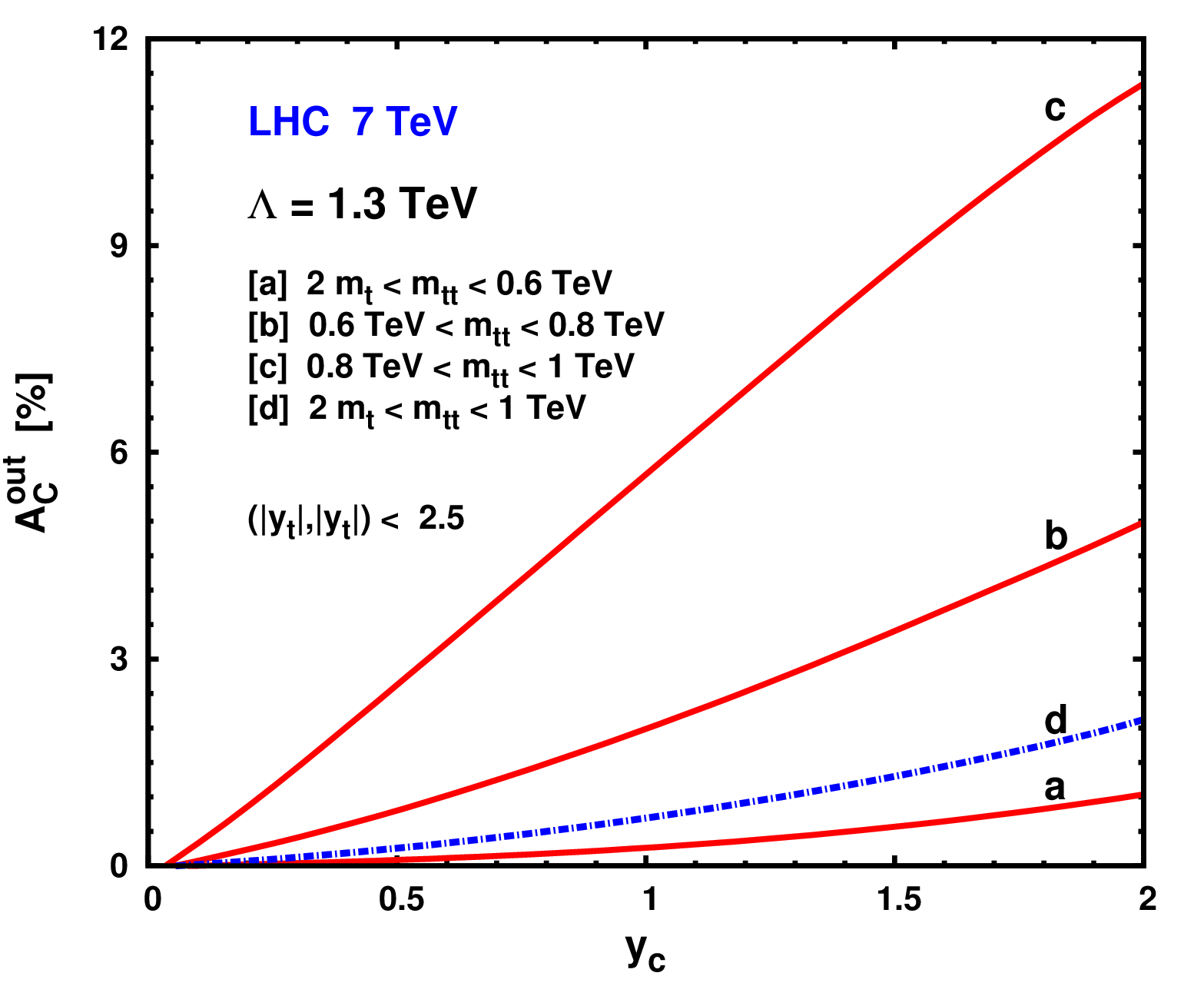}
\includegraphics[width=0.40\textwidth, angle=-90]{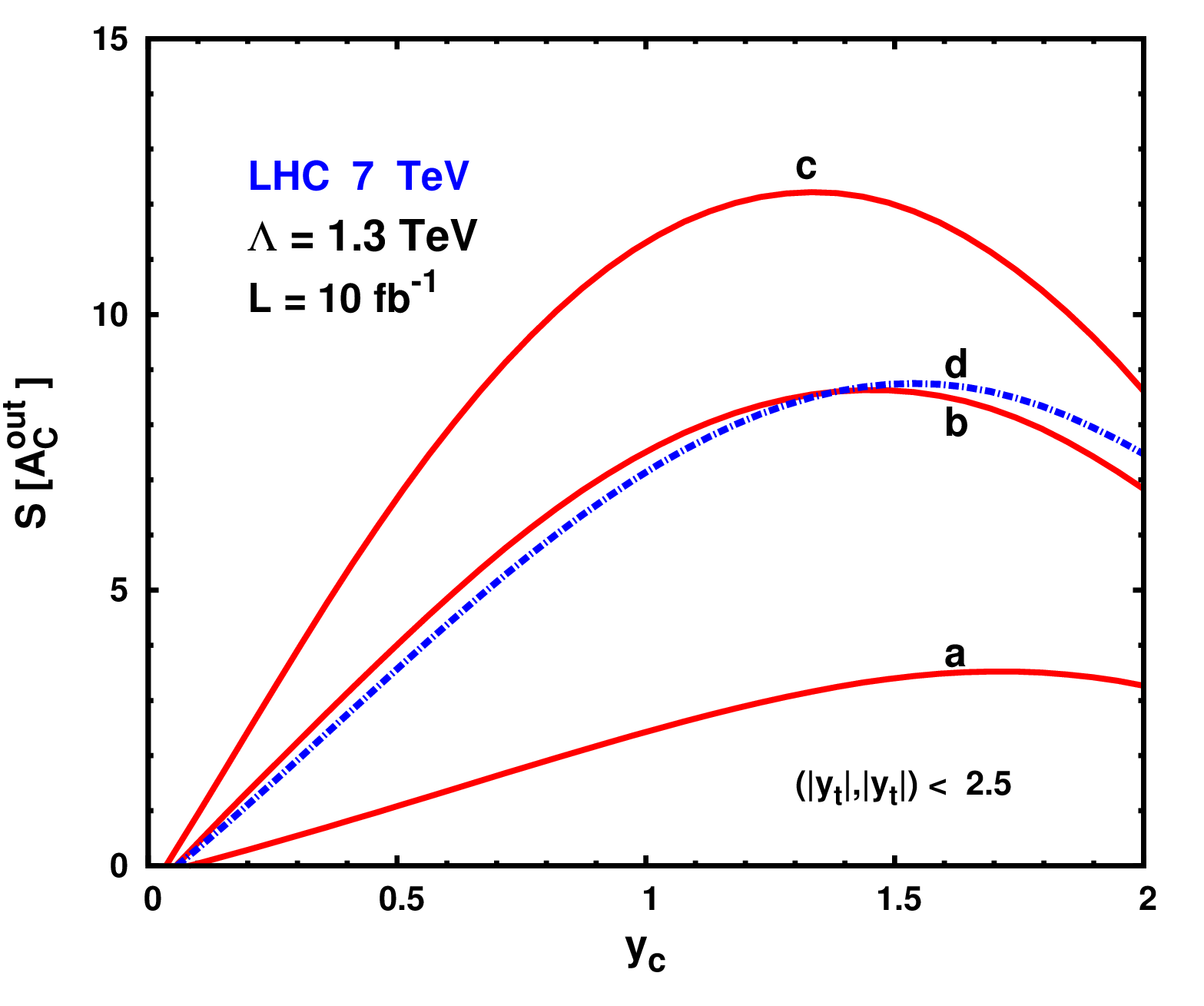}
\vspace{-0.3cm}
\caption{The $t\bar t$ charge asymmetry $A_{C}^{out}(y_c)$
in percentage  (left plots)
and corresponding statistical significance $S[A_C^{out}]$ (right plots)
at LHC with $pp$ center of mass energy $\sqrt{S}$ = 7 TeV
and integrated luminosity $L=10~{\rm fb}^{-1}$, with $m_t=172$ GeV,
as a function of the cuts on the $t$- and $\bar t $-quark rapidity $y_c$
(in the lab frame),
for several regions ([a-d]) of $t\bar t$ invariant mass $m_{tt}$.
The upper constraints on rapidities $(|y_t|, |y_{\bar t}|) < 2.5$ is imposed.
Up and down plots correspond to the scale $\Lambda=1$ TeV
and $\Lambda=1.3$ TeV,  respectively.}
\label{fig:ACout}
\end{center}
\end{figure*}

\begin{figure*}[t]
\begin{center}
\includegraphics[width=0.40\textwidth, angle=-90]{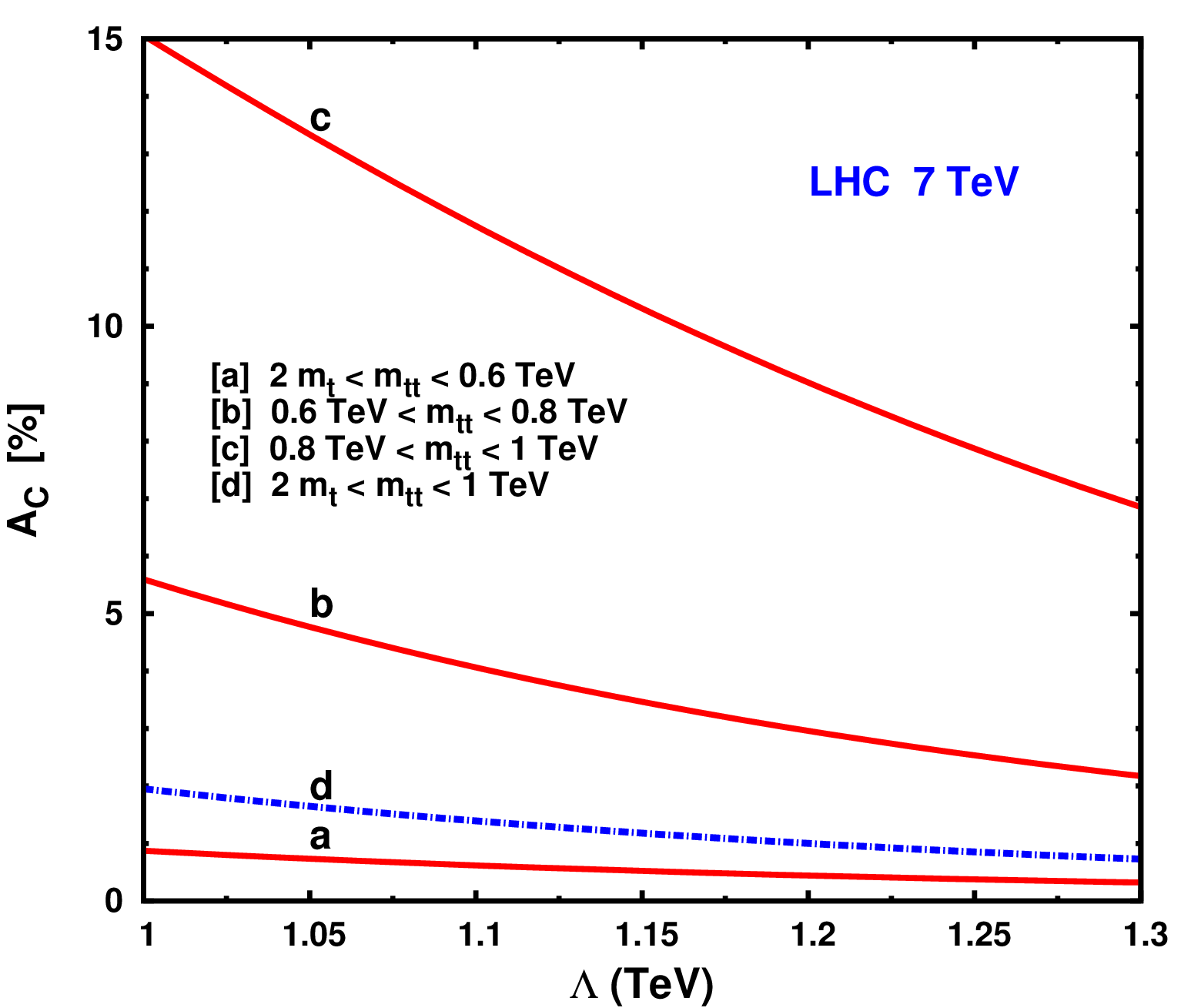}
\includegraphics[width=0.40\textwidth, angle=-90]{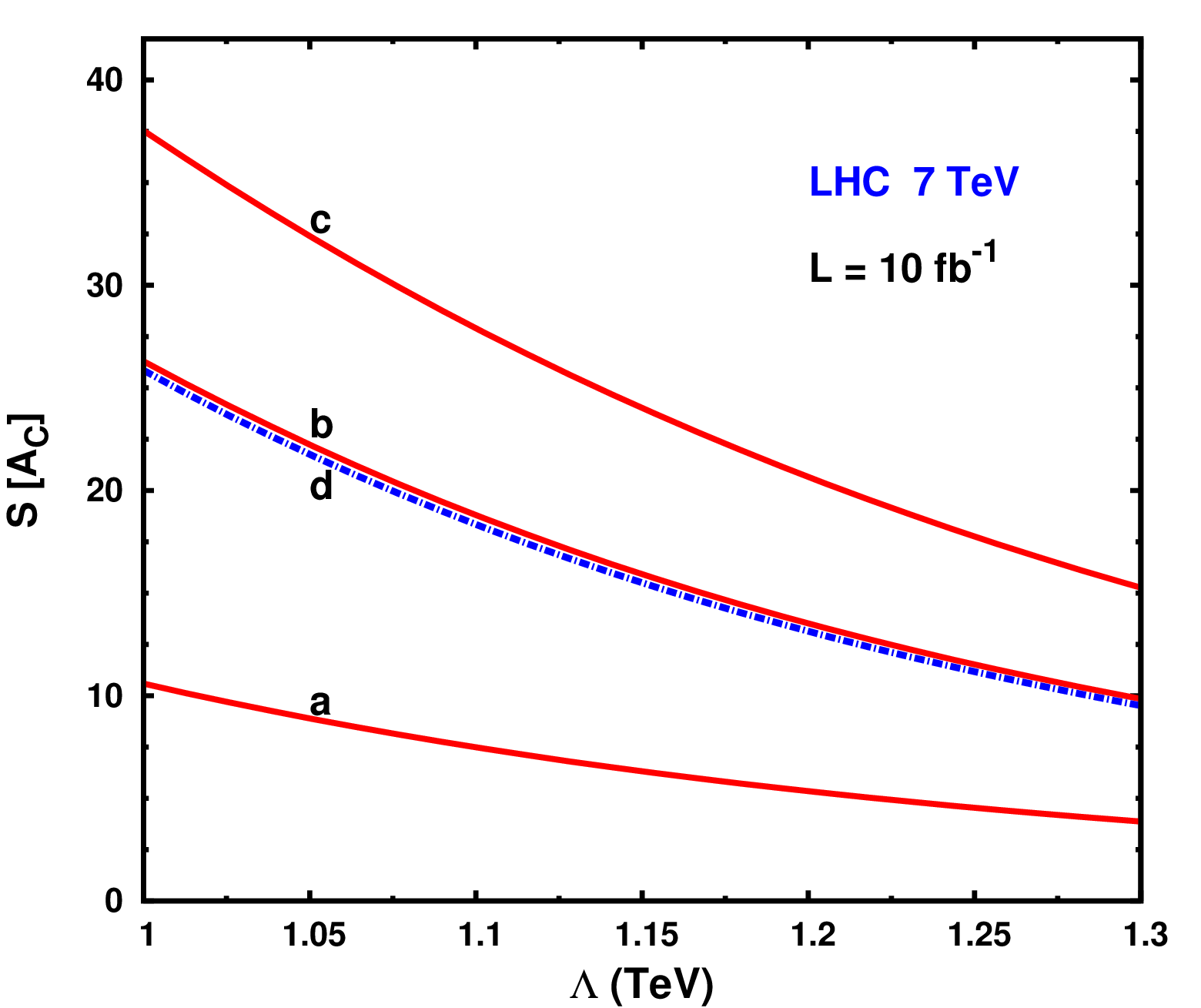}
\vspace{-0.3cm}
\caption{The cut-independent  $t\bar t$ charge asymmetry $A_{C}$
in percentage  (left plots)
and corresponding statistical significance $S[A_C]$ (right plots)
at LHC with $pp$ center of mass energy $\sqrt{S}$ = 7 TeV
and integrated luminosity $L=10~{\rm fb}^{-1}$, with $m_t=172$ GeV,
as a function of the scale $\Lambda$ in TeV,
for several regions ([a-d]) of $t\bar t$ invariant mass $m_{tt}$.}
\label{fig:AC}
\end{center}
\end{figure*}

\begin{figure*}[t]
\begin{center}
\includegraphics[width=0.40\textwidth, angle=-90]{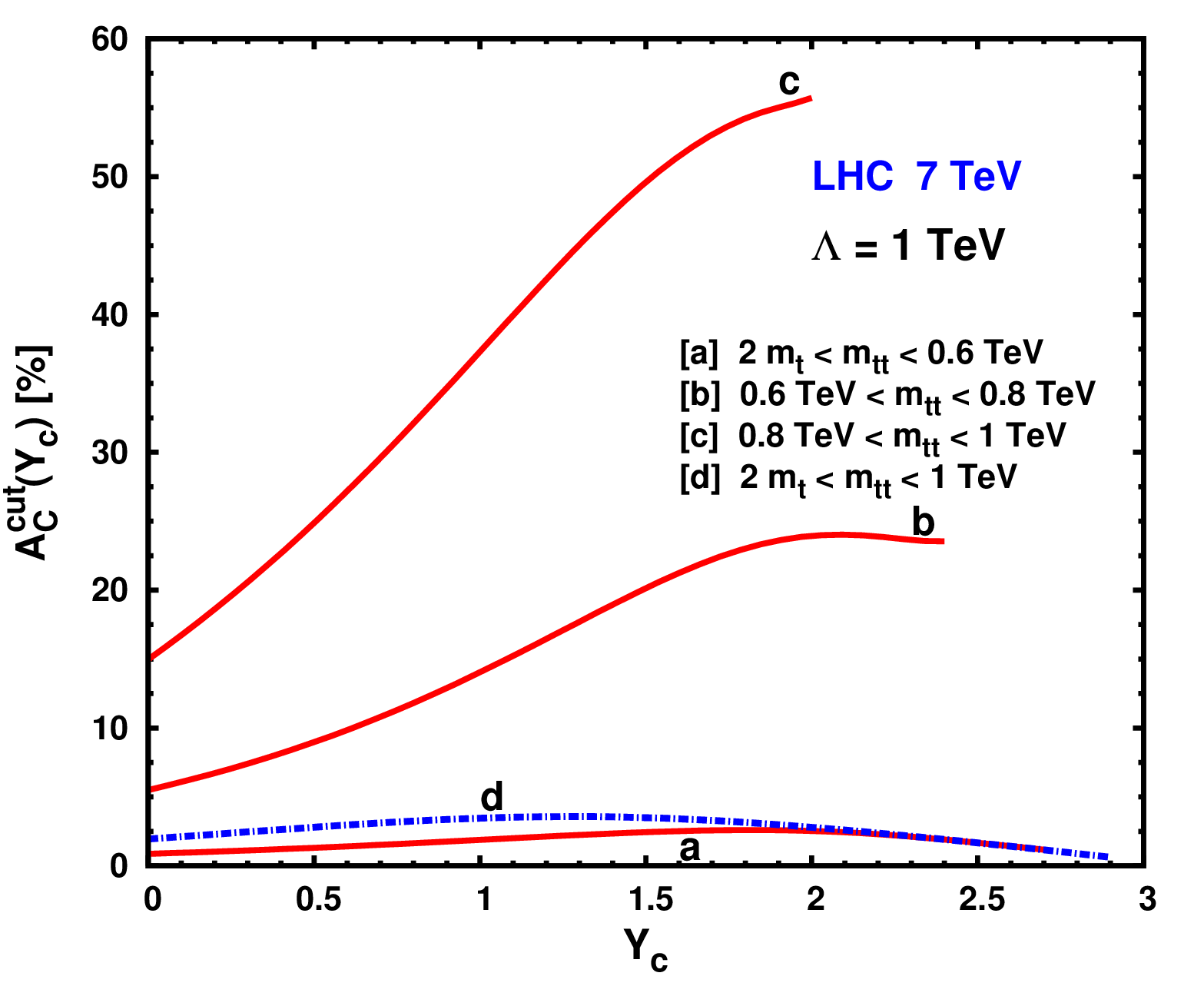}
\includegraphics[width=0.40\textwidth, angle=-90]{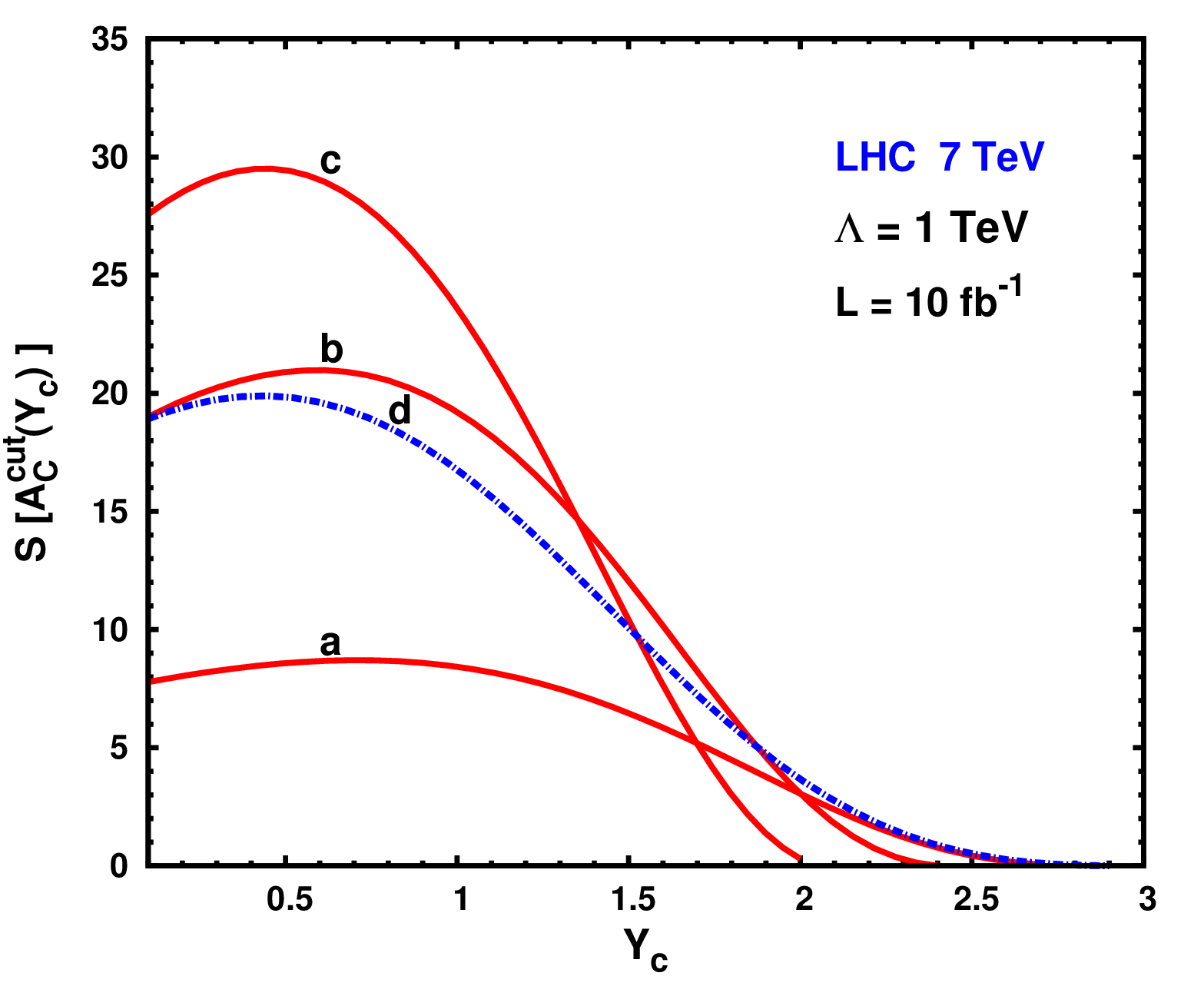}
\includegraphics[width=0.40\textwidth, angle=-90]{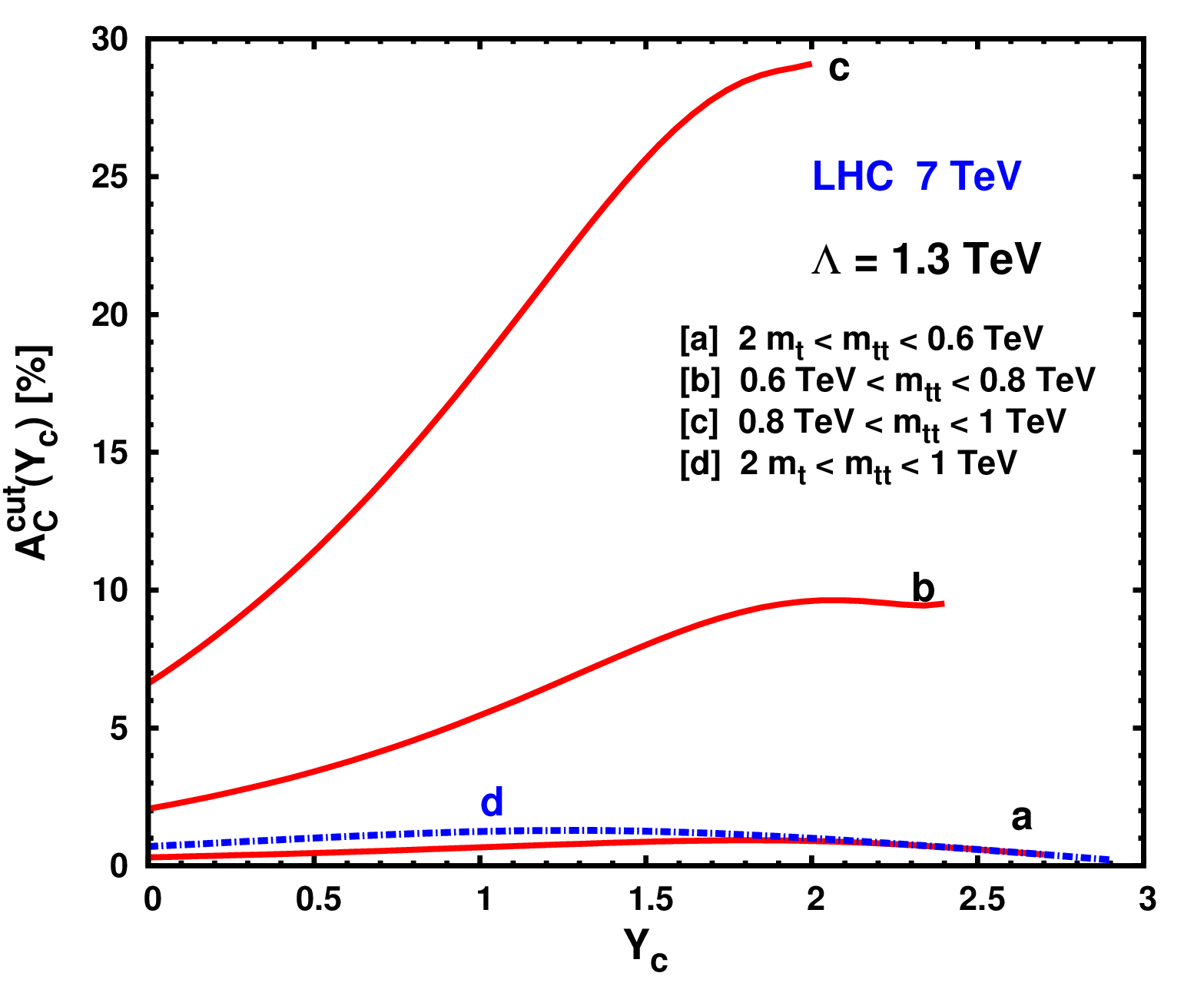}
\includegraphics[width=0.40\textwidth, angle=-90]{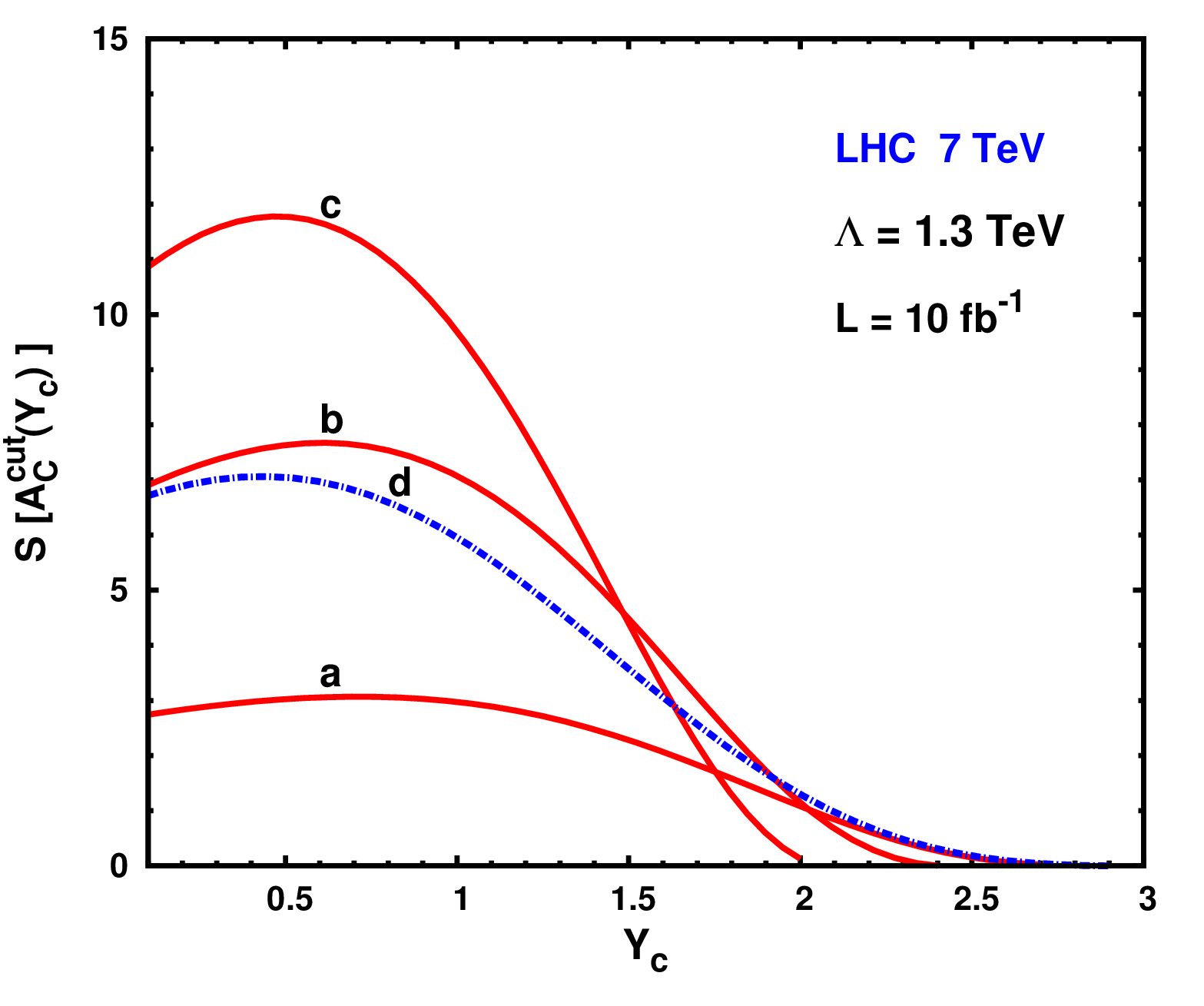}
\vspace{-0.3cm}
\caption{Cut-dependent pair charge asymmetry $A^{\rm cut}_{C}(Y_c)$
in percentage  (left plots)
and corresponding statistical significance $S[A_C^{\rm cut}(Y_c)]$ (right plots)
at LHC with $pp$ center of mass energy $\sqrt{S}$ = 7 TeV
and integrated luminosity $L=10~{\rm fb}^{-1}$, with $m_t=172$ GeV,
as a function of the cuts on the mean rapidity
$|Y|> Y_c$ (in the lab frame),
for several regions ([a-d]) of $t\bar t$ invariant mass $m_{tt}$.
Up and down plots correspond to the scale $\Lambda=1$ TeV
and $\Lambda=1.3$ TeV,  respectively.}
\label{fig:ACcut}
\end{center}
\end{figure*}

\begin{figure*}[t]
\begin{center}
\includegraphics[width=0.40\textwidth, angle=-90]{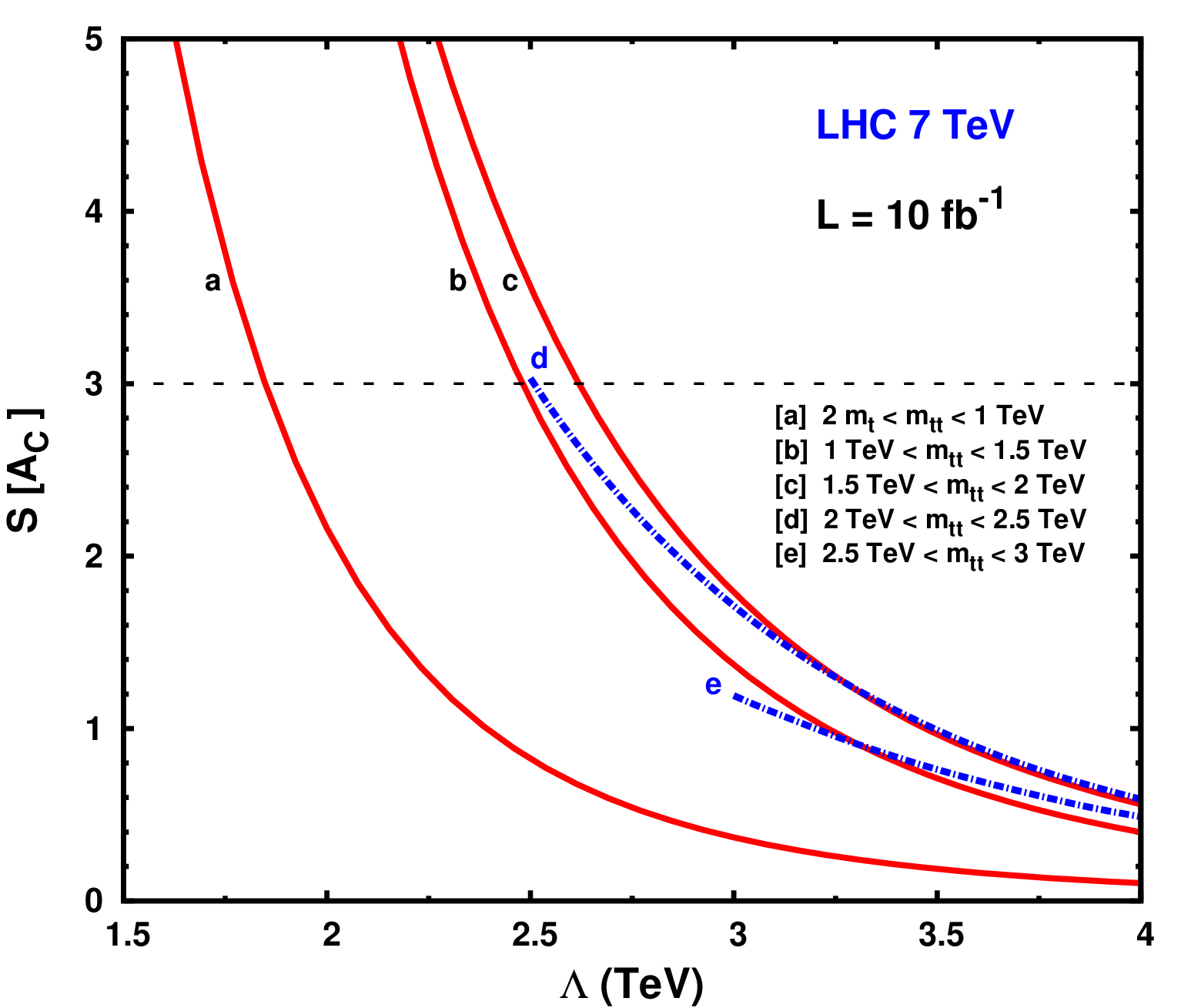}
\includegraphics[width=0.40\textwidth, angle=-90]{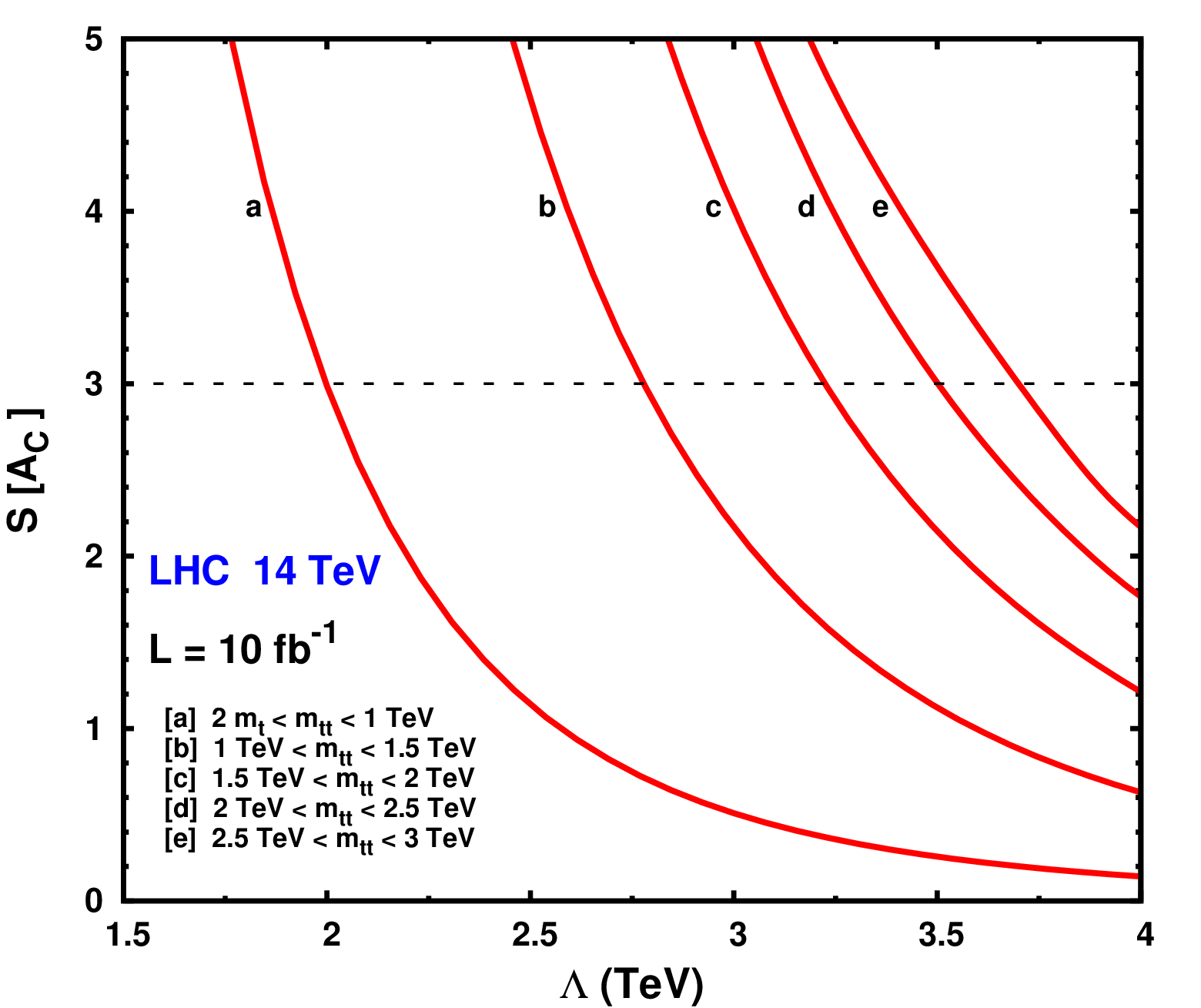}
\vspace{-0.3cm}
\caption{Statistical significance $S[A_C]$ of the cut-independent $t\bar t$
charge asymmetry $A_C$
at LHC with $pp$ center of mass energies $\sqrt{S}$ = 7 TeV (left plots)
and $\sqrt{S}$ = 14 TeV (right plot)
with  integrated luminosity $L=10~{\rm fb}^{-1}$, with $m_t=172$ GeV,
as a function of the scale $\Lambda$ in TeV,
for several regions ([a-e]) of $t\bar t$ invariant mass $m_{tt}$.}
\label{fig:S}
\end{center}
\end{figure*}

Our numerical results for the $g_A$ contributions to the charge asymmetries,
defined in Eqs.~(\ref{CAin})-(\ref{CAcut}), together with
their corresponding statistical significances,
are shown in Figs.~\ref{fig:ACin}-\ref{fig:S}.

Regarding the statistical significance $S[A_C]$ for a generic
charge asymmetry definition, we have used the approximated
relation \cite{Ferrario:2008wm} 
\bea S[A_C] \simeq  A_C \sqrt{L\,
\sigma^{\rm NP+SM}}\, , 
\label{signf}
\eea 
where $L$ stands for the integrated
luminosity and $\sigma^{\rm NP+SM}$ is the total cross section
including SM and NP contribution. In $\sigma^{\rm NP+SM}$  we have
used the LO cross sections multiplied by the rescaling factor $K$.
This $K$ factor is obtained by simply rescaling the total cross
sections evaluated at the LO in QCD 
to its value corrected at the next-to-next-to-leading
order (NNLO) in QCD \cite{Kidonakis:2009mx,Kidonakis:2011jg}. 
Although the difference is of order of few
percent, we have used two separate rescaling factors for the LHC
center of mass energies corresponding $\sqrt{S}=7$~TeV and
$\sqrt{S}=14$~TeV. Moreover, we have assumed a universal $K$
factor for different kinematic regions. All plots of
significances, correspond to an integrated luminosity of $L=10~{\rm fb}^{-1}$.
Notice that 
the significance in Eq.(\ref{signf}) is a simple theoretical
estimation of the true one, since it does not take into account efficiencies, 
acceptance, resolution, and systematics. 

In the left side plots of Fig.~\ref{fig:ACin} we show the values
of the {\bf in} cut-dependent charge
asymmetry $A_C^{\rm in}$, as a function of the rapidity cut $y_c$ in the
range $0.1 < y_c < 3$, for four different kinematic regions [a-d], corresponding
to $m_{tt}$ cuts in the following ranges:
\bea
[a] &=& 2 m_t < m_{tt} < 0.6~  {\rm TeV}, \label{ranges} \\
\nonumber
[b] &=& 0.6~{\rm TeV} < m_{tt} < 0.8~ {\rm TeV},\\
\nonumber
[c] &=& 0.8~{\rm TeV} < m_{tt} < 1~ {\rm TeV},\\
\nonumber
[d] &=& 2 m_t < m_{tt} < 1~ {\rm TeV},
\eea
and for $\Lambda=1$ TeV (top left-plot) and $\Lambda=1.3$ TeV (down left-plot).
We have used the same definition of
$m_{tt}$ ranges in Eq.~(\ref{ranges}) in all
Figs.~\ref{fig:ACin}-\ref{fig:ACcut}. From now on, if not specified,
we will refer to the definition of [a]-[d] $m_{tt}$ ranges according
to Eq.~(\ref{ranges}).

The SM contribution to the top-quark charge asymmetries has been
recently updated in Refs.~\cite{Kuhn:2011ri,Hollik:2011ps}, 
where the one-loop EW
corrections have been included, amounting roughly to 1.1\% ~\cite{Kuhn:2011ri}.
The SM predicts always a positive sign for all the charge asymmetries
defined in Eqs.~(\ref{CAin})-(\ref{CAcut}).

One general prediction of our scenario is that the effective
axial-vector gluon coupling contributes to all top-quark charge
asymmetries with the same sign as the SM one. From the results of
Ref.~\cite{Kuhn:2011ri} we can see that the magnitude of the SM
contribution to $A_C^{\rm in}(y_c)$  is quite small, being below
1\% in the full range of $y_c$.\footnote{ Notice that a direct
comparison between our results and the SM predictions in
\cite{Kuhn:2011ri} is actually misleading, since in
\cite{Kuhn:2011ri} $m_{tt}$ has been integrated over all the
allowed kinematic range. In our scenario, this is not possible,
being the low energy approximation for the $g_A$ form factor only
valid for values of $m_{tt} < \Lambda$. Nevertheless, due to the
fact that the SM charge-asymmetry is weakly dependent on $m_{tt}$,
we expect that results in Ref.~\cite{Kuhn:2011ri} will not
dramatically change if restricted, for instance, to the kinematic
region of $2 m_t < m_{tt} < 1 {\rm TeV}$, where a direct
comparison with our results is possible.} On the other hand, in
our scenario $A_C^{\rm in}(y_c)$ can be substantially enhanced
well above the few percent level, depending on the $m_{tt}$
integrated regions.

In the right plots of Fig.~\ref{fig:ACin}, we show the
values for the corresponding significance $S[A_C^{\rm in}]|$ as a function of
$y_c$ in the case of $\Lambda=1$ TeV (up plot) and $\Lambda=1.3$ TeV (down
plot).
From these results we can see that its maximum value is
reached for rapidity cuts around $y_c\sim 0.7$, roughly in all
$m_{tt}$ ranges [a]-[d].
In particular, at $y_c=0.7$
and for $L=10~ {\rm fb}^{-1}$, we get
$S[A_C^{\rm in}]|_{\rm max}\simeq (30,19,16,6)$ for $\Lambda=$ 1 TeV
and $S[A_C^{\rm in}]|_{\rm max} \simeq (12,7,6,2)$
for $\Lambda=$ 1.3 TeV, corresponding to $m_{tt}$=([c],[b],[d],[a]),
respectively. For these ranges, the values of the charge asymmetries in percentage
are $A_C^{\rm in} [\%]=(14,4.4,1.3,0.5)$ and
$A_C^{\rm in} [\%]=(6.4,1.6,0.5,0.2)$ for  $\Lambda=$ 1 TeV and
$\Lambda=$ 1.3 TeV, respectively.

In Fig.~\ref{fig:ACout} we report the results for the $A_C^{\rm out}(y_c)$ asymmetry
(left plots) and corresponding significance (right plots), as a function
of the rapidity cuts $y_c$, obtained by also imposing an
upper limit on top and anti-top rapidities, namely
$|y_t|<2.5$, $|y_{\bar t}|<2.5$.
According to these results, we can see
that the maximum significance is reached in the range $y_c=[1.4-1.6]$,
with $S[A_C^{\rm in}]|_{\rm max}\simeq (33,24,25,10)$
for $\Lambda=$ 1 TeV and $S[A_C^{\rm in}]|_{\rm max} \simeq (12,9,9,3)$
for $\Lambda=$ 1.3 TeV, corresponding to
the $m_{tt}$ ranges ([c],[b],[d],[a]), respectively.
For the corresponding charge asymmetry in percentage, evaluated
for instance at $y_c=1.5$,
we get $A_C^{\rm out} [\%]=(21,9,3.6,1.6)$ and
$A_C^{\rm out} [\%]=(8.8,3.4,1.3,0.6)$ in the ranges ([c],[b],[d],[a]),
for  $\Lambda=$ 1 TeV and $\Lambda=$ 1.3 TeV, respectively.
From these results we can see that $A_C^{\rm out}$ is more sensitive  than
$A_C^{\rm in}$ to the axial-vector coupling and at $y_c=1.5$ can provide a
larger asymmetry with better significance with respect to $A_C^{\rm in}$.

In Fig.~\ref{fig:AC} we plot the
cut-independent  $t\bar t$ charge asymmetry $A_{C}$ (left plot)
and corresponding significance (right plot) as a function of the
scale $\Lambda$ in the range [1-1.3] TeV.
This is the definition adopted by the ATLAS 
\cite{ATLAS-CONF-2011-106} and CMS~\cite{CMS-PAS-TOP-11-014,:2011hk}  
collaborations to measure the top-quark charge asymmetry at LHC. As shown from
these results, the $g_A$ contribution to $A_C$ is quite large when
measured at high $m_{tt}$ masses close to the value of the scale
$\Lambda$. In particular, for the range [c], $A_C$ could be of
order of 15\% and 7\%, for a scale $\Lambda=1$ TeV and
$\Lambda=1.3$ TeV respectively, while its corresponding
significance can reach values of 37 and 15, respectively. On the
other hand, when integrated over a large $m_{tt}$ range, see for
instance the curve relative to [b] range, $A_{C}$ turns out to be
quite small (below 2 \%) and comparable (although a bit larger) to
the SM result \cite{Kuhn:2011ri}.

Finally, in Fig.~\ref{fig:ACcut} we show the cut-dependent pair charge asymmetry
$A^{\rm cut}_{C}(Y_c)$ as defined in Eq.~(\ref{CAcut})
(left plots) and corresponding significance
(right plots), as a function of  the cuts $Y_c$ on the mean rapidity
$Y=(y_t+y_{\bar t})/2$,
for the representative values of $\Lambda=1, 1.3$ TeV.
As we can see from these results, $A^{\rm cut}_{C}(Y_c)$ turns out
to be 
the best sensitive probe of our scenario. In particular, at $\Lambda=1$ TeV,
for $m_{tt}$ in the range [c], the value of $A^{\rm cut}_{C}(Y_c)$ can vary,
as a function of $Y_c$, from 15\%  up to 55\% for $\Lambda=1$ TeV
and from 7\% up to 29\% for for $\Lambda=1.3$ TeV.
The value of $A^{\rm cut}_{C}(Y_c)$ in the range [c] evaluated for $Y_c=0.3$,
where its significance is maximized, is of order of 20\% and 10\% for
$\Lambda=1$ TeV and
$\Lambda=1.3$ TeV, respectively. However, even at $Y_c=1.5$
the cut-dependent pair charge asymmetry is still quite large.
For $\Lambda=1$ TeV, $A^{\rm cut}_{C}(1.5)$
is still of order  50\% and 20\% if integrated in the
ranges [c] and [b] respectively,
while the corresponding significances are still above 10.
However, for $\Lambda=1.3$ TeV, the value of $A^{\rm cut}_{C}(1.5)$
drops down to 25\% and 7\% for $m_{tt}$
ranges [c] and [b] respectively, with a corresponding
significance of order 5.

In Fig.~\ref{fig:S} we analyzed the statistical significance of
the cut-independent charge asymmetry $A_C$ versus the scale $\Lambda$ in the
range of $\Lambda$=[1.5-4] TeV and for several ranges of $m_{tt}$ as indicated
in the figure.
In particular, the left and right plots correspond
to the LHC center of mass energies of $\sqrt{S}=$ 7 TeV and $\sqrt{S}=$ 14
TeV, respectively.
Regarding the choice of ranges of $m_{tt}$,
we adopted the criteria that the maximum value of the $m_{tt}$ satisfies the
condition
$m_{tt}^{\rm max} \leq \Lambda$ as required by the validity range of
the low energy limit in the $g_A$
form factor.  Assuming that no sensitive deviations
from SM predictions on the charge asymmetry $A_C$ will be observed,
from results in Fig.~\ref{fig:S} one can derive lower bounds on
the scale $\Lambda$. For example,
by requiring that $S[A_C] < 3$, we get for $L=10~{\rm fb}^{-1}$,
the following (strongest) lower bounds:
\begin{itemize}
\item
$\Lambda  > 2.6~ {\rm TeV}$, for
$m_{tt}\in$ [1.5-2] TeV at LHC 7 TeV;
\item
$\Lambda > 3.7$ TeV, for $m_{tt}\in$ [2.5-3] TeV at LHC 14 TeV\,.
\end{itemize}
As can be seen from the dashed (blue) curves in the left plot of
Fig.~\ref{fig:S}, for LHC 7 TeV there is not any advantage, concerning the
lower bounds of $\Lambda$, in going to higher bin ranges in $m_{tt} > 2$ TeV,
due to the loss of statistics.

\begin{figure*}[t]
\begin{center}
\includegraphics[width=0.40\textwidth, angle=-90]{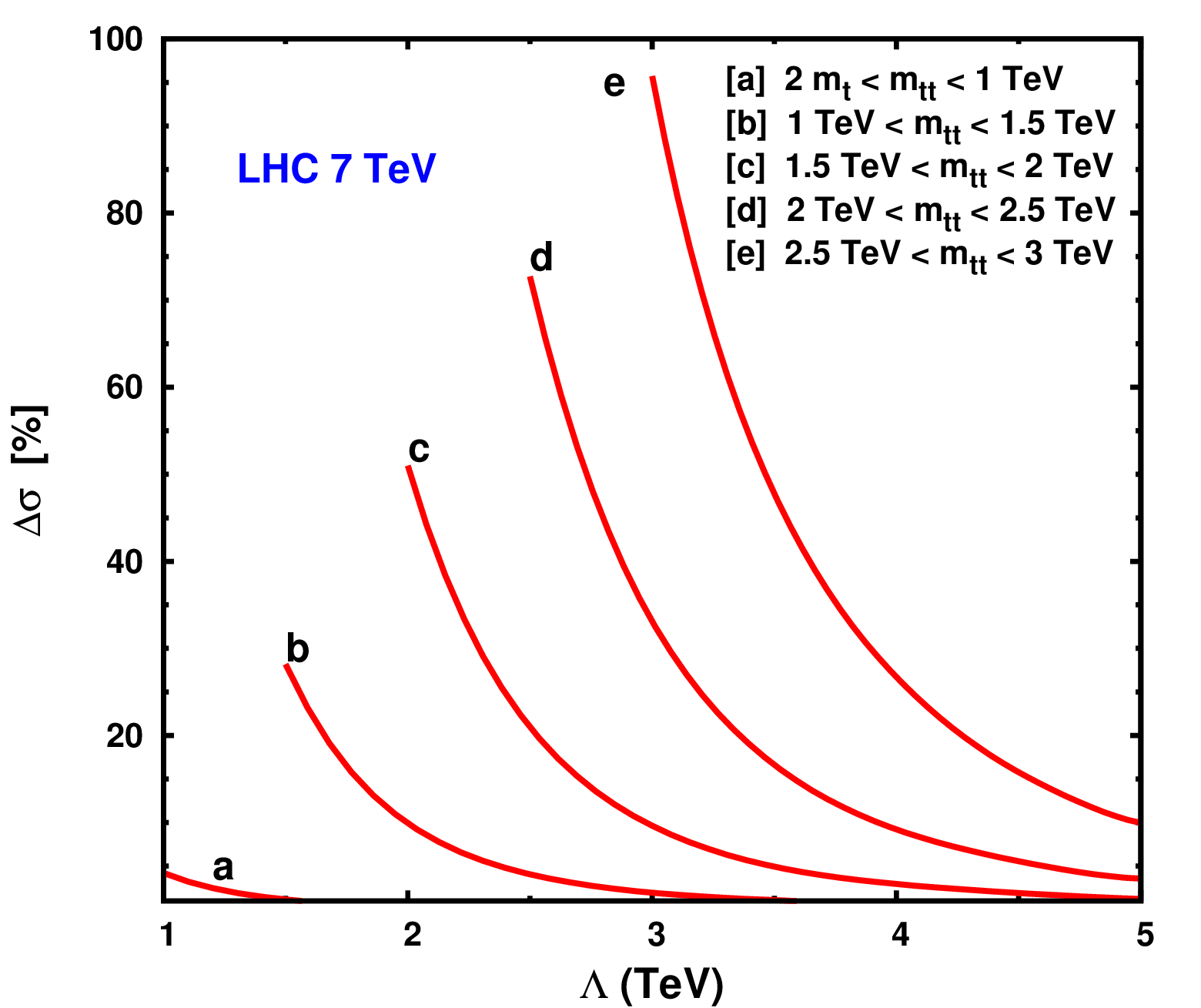}
\includegraphics[width=0.40\textwidth, angle=-90]{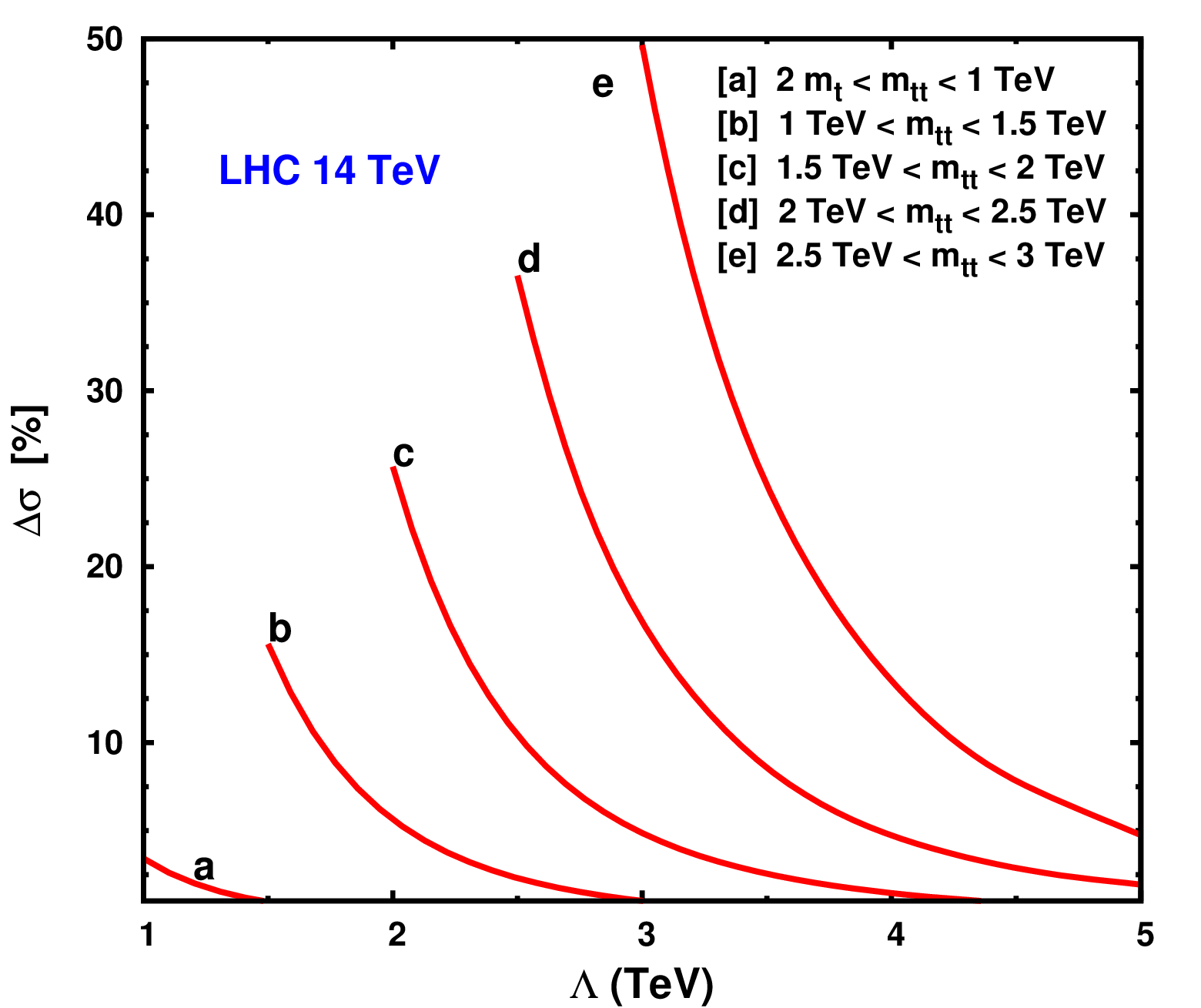}
\vspace{-0.3cm}
\caption{Percentage variation $\Delta \sigma$
of the total cross section for $pp \to t\bar t$
at LHC with $pp$ center of mass energies $\sqrt{S}$ = 7 TeV (left plots)
and $\sqrt{S}$ = 14 TeV (right plot) for $m_t=172$ GeV,
as a function of the scale $\Lambda$ in TeV,
for several regions ([a-e]) of $t\bar t$ invariant mass $m_{tt}$.}
\label{fig:deltasigma}
\end{center}
\end{figure*}

In Fig.~\ref{fig:deltasigma} we plot the percentage variation $\Delta \sigma$ of the
total cross section for $pp \to t \bar t$ at LHC for LHC 7 TeV (left)
and LHC 14 TeV (right), as a function of $\Lambda$
from 1 TeV to 5 TeV, and for several ranges of $m_{tt}$.
The $\Delta \sigma$ is defined after Eq.~(\ref{asym})
and cross sections have been evaluated at the LO in QCD. 
The picture that emerges
from these results is clear. If we analyze ranges of
$m_{tt}$ below 1 TeV scale, see the curve corresponding to the [a] range,
the expected percentage variation in the total cross section is quite
small, being below 5\% in both energy ranges at LHC 7 TeV and 14 TeV.
This is due to the fact that requiring $m_{tt}$ to be below 1 TeV,
the $g_A$ contribution at low energy is still largely
screened by the gluon-gluon fusion mechanism,
which is the dominant mechanism of top-quark pair production
at LHC. Clearly, by increasing the $m_{tt}$ mass range the NP effect could be
amplified and larger deviations could be observed in the cross sections,
due to the quark-antiquark production mechanism.
These results could also be used to set lower bounds on the scale $\Lambda$
by requiring that no excess in the total cross section is observed with
respect to the SM predictions. For instance, by limiting the deviations
on total cross section below 20 \%, one can see from the curve [e] that a
lower bound $\Lambda >4 $ TeV could be obtained at LHC 7 TeV. Clearly,
by increasing the $m_{tt}$ range, the statistical error on the cross section
also increases, and a more accurate analysis, which is going beyond
the purpose of the present paper, would be necessary.

\section{Conclusions}
We studied the gluon effective axial-vector coupling induced top charge asymmetries at the LHC.
We compared rapidity cut-dependent and independent asymmetries and showed that the former are
more sensitive to NP than the latter. We also studied the asymmetries and variations of
total $t\bar t$ cross sections at different invariant masses of the $t\bar t$ system and showed that
it would be necessary to measure those quantities as functions of $m_{tt}$ at the LHC.
If this is done, 7~TeV LHC has enough sensitivity either to confirm the Tevatron top charge
  asymmetry anomaly or to rule it out in the context of considered NP scenario.
  In the latter case the LHC is able to put stringent constraint on the NP scale $\Lambda.$

\section*{Acknowledgment}
We thank T. Chwalek, A. Giammanco, T. Peiffer,
G. Rodrigo, and J. Wagner-Kuhr for several communications.
This work was supported by the ESF grants  8090,  MTT59, MTT60, JD164,  by the recurrent financing SF0690030s09 project
and by  the European Union through the European Regional Development Fund.

\bibliography{revised_prd.bib}

\end{document}